\documentclass[fleqn,usenatbib]{mnras}

\usepackage[T1]{fontenc}
\usepackage{color}
\DeclareRobustCommand{\VAN}[3]{#2}
\let\VANthebibliography\thebibliography
\def\thebibliography{\DeclareRobustCommand{\VAN}[3]{##3}\VANthebibliography}

\usepackage{geometry}
\usepackage{marginnote}
\definecolor{Red}{rgb}{0.743,0,0}
\definecolor{Blue}{rgb}{0.25,.41,.88}
\definecolor{Green}{rgb}{0,0.5,0}

\usepackage{graphicx}	
\usepackage{amsmath}	
\usepackage{amssymb}	
\usepackage{gensymb}

\usepackage{newtxtext,newtxmath}

\def\sgra{Sgr~A$^*$}

\newcommand\lsim{\mathrel{\rlap{\lower4pt\hbox{\hskip1pt$\sim$}}
    \raise1pt\hbox{$<$}}}
\newcommand\gsim{\mathrel{\rlap{\lower4pt\hbox{\hskip1pt$\sim$}}
    \raise1pt\hbox{$>$}}}

\title[Constraints on the origins of hypervelocity stars]{Constraints on the origins of hypervelocity stars: velocity distribution, mergers and star-formation history}

\author[Generozov \& Perets]{
Aleksey Generozov,$^{1}$\thanks{E-mail: al.generozov@campus.technion.ac.il}
Hagai B. Perets,$^{1}$
\\
$^{1}$Physics Department, Technion — Israel Institute of Technology, Haifa 3200003, Israel\\
}

\date{Accepted XXX. Received YYY; in original form ZZZ}

\pubyear{2022}

\begin{document}
\label{firstpage}
\pagerange{\pageref{firstpage}--\pageref{lastpage}}
\maketitle

\begin{abstract}
In recent years surveys have identified several dozen B stars in the Milky Way halo moving faster than the local escape speed. 
The origin of most of these hypervelocity stars (HVSs) is still poorly constrained. Here we show that the velocity distribution, and in particular the deficiency in >700 km s$^{-1}$ HVSs is inconsistent with binary disruptions by the massive black hole (MBH) in the Galactic Centre. This conclusion holds in the full and empty loss cone regime, and for secular instabilities in eccentric disks. Accounting for multiple close encounters between binaries and the MBH, does not qualitatively change the results. Moreover, there is no observed counterpart population in the Galactic Centre that is consistent with the HVSs. The star-formation history could be tuned explain the HVS velocity distribution, but this tuning would produce a mismatch with the observed HVS flight times. Frequent stellar collisions of the binary components due to interactions with the MBH do not significantly impact the velocity distribution in the Galactic halo. Such  collisions, however,  can leave observable remnants in the Galactic Centre, and potentially explain the origins of G2-like dust clouds. 
\end{abstract}



\section{Introduction}

\cite{hills1988} first proposed tidal disruption of binaries by a supermassive black hole can populate the Galactic halo with hypervelocity stars -- stars with velocities exceeding the Galactic escape speed. Hypervelocity stars ejected from the Galactic Centre will follow a radial trajectory, modulo perturbations from a non-spherical potential. Thus, these stars probe the shape of the Galactic potential  \citep{gnedin+2005,yu&madau2007,perets+2009_galPot,contigiani+2019}.

After the serendipitous discovery of the first hypervelocity star, HVS-1, by \citet{warren_brown+2005}, a few dozen hypervelocity star candidates have been identified.
The history of these discoveries is reviewed in \citet{warren_brown2015}, and we briefly summarize it here.
A few more candidates were found serendipitously immediately after HVS-1 \citep{hirsch+2005, edelmann+2005}. 
These discoveries motivated the targeted hypervelocity star survey (\citealt{warren_brown+2014} and the references therein), which has identified the largest number of hypervelocity star candidates to date using radial velocities. Other candidates have been identified using a combination of radial and tangential velocities \citep{heber+2008, zheng+2014} or primarily by tangential velocities  \citep{tillich+2009,irrgang+2010,tillich+2011,li_hvs+2012, pereira+2012, pereira+2013,palladino+2014, zhong_hvs+2014}. The candidates identified by radial velocities are early-type (mostly B) stars, while the candidates identified by tangential velocities are mostly late-type stars. Using Gaia DR2 data \citet{boubert+2018} found most of the tangential velocity stars are actually bound to the Galaxy, and consequently ruled out all but one of the historical late-type hypervelocity candidates. More recently, \citet{marchetti2021} identified 17 late-type stars in Gaia EDR3 that are likely unbound.

There are many proposed explanations for the observed hypervelocity star candidates. They may be produced by binaries pushed to disruption by ordinary stars, massive perturbers \citep{perets+2007}, spiral arms \citep{hamers&perets2017}, or secular gravitational instabilities in a stellar disc \citep{madigan+2009}. Stars may also be ejected from the Milky Way's Galactic Centre  by stellar mass black hole kicks \citep{yu&tremaine2003,oleary&loeb2008} or IMBH inspirals \citep{yu&tremaine2003,baumgardt+2006,levin2007}.
Cluster infall can also produce hypervelocity stars \citep{fragione+2017}.

Finally, they may originate from outside the Galactic Centre entirely:\footnote{We note that in some papers the term ``hypervelocity star'' is reserved for unbound stars ejected from the Galactic Centre. Other unbound stars are referred to as hyper-runaways (see \citealt{boubert+2017} and the references therein).} from the LMC \citep{edelmann+2005,gualandris&portegies-zwart2007,boubert+2017,erkal+2019}, supernova kicks \citep{blaauw1961}, disrupting dwarf galaxies \citep{abadi+2009}, hyper-runaways from a dense cluster \citep{poveda+1967,leonard1991,gvaramadze+2009,perets&subr2012}, or disruption of binaries by IMBHs in star clusters \citep{fragione&gualandris2019}. Some of these mechanisms can also operate in the Galactic centre. For example, supernova kicks in the Galactic Centre may produce large ejection velocities than in other environments due to the large stellar velocities there \citep{zubovas+2013}.

The origins of some candidates can be constrained by integrating their orbits backwards in time in a model Galactic potential. In particular, a Galactic Centre origin has been ruled out for all existing late-type candidates \citep{marchetti2021}. However, due to astrometric uncertainties, the origin of many candidates cannot be directly constrained. Instead spatial anisotropies \citep{warren_brown+2009} and rate arguments \citep{perets2009,perets&gualandris2010} may be used to distinguish or rule out channels. In fact, rate arguments suggest that IMBH inspirals and stellar mass black holes cannot fully account for the observed population. As discussed below, the velocity distribution itself can also be a powerful discriminant.

With the notable exception of S5-HVS1, a 1700 km s$^{-1}$ star ejected from the Galactic Centre $\sim$5 Myr ago \citep{koposov+2019}, the observed sample of the hypervelocity star candidates does not include stars with velocities above 700 km s$^{-1}$. In contrast, previous work has estimated $\sim 40\%$ of hypervelocity stars ejected by binary disruptions should exceed this speed for a log-uniform semimajor axis distribution and isotropic loss-cone models \citep{rossi+2014}. A log-normal semimajor axis distribution, like that of observed solar-type binaries, provides a better match to the observed velocities \citep{sesana+2007_hvs}. However, this distribution is not a good fit to early-type binaries relevant to the observed hypervelocity star candidates \citep{moe+2017}. Additionally, multiple encounters between binaries and the central black hole can truncate the observed velocity distribution at $\sim$1000 km s$^{-1}$ by causing binaries to expand in semimajor axis prior to disruption and by causing disruptions to occur further away from the MBH \citep{zhang+2010_hvs, zhang+2013_hvs}.
This mechanism will only work if the progenitor binaries start very close to MBH, and experience many close encounters prior to disruption.  \citet{madigan+2009} and \citet{generozov&madigan2020} describe how  binaries in young disc(s) at small galactocentric radii \citep{levin&beloborodov03,paumard+2006} can be efficiently torqued to disruption by a secular gravitational instability. However, in this case, the binaries do not random walk to disruption (as assumed in \citealt{zhang+2010_hvs}), but experience coherent changes in angular momentum over many orbits around the Galactic Centre, reducing the number of close encounters with the MBH.

Here, we predict the velocity distribution of hypervelocity stars for different source binary populations: (i) A spherical population of binaries within the central few parsecs of the Galaxy, (ii) Binaries from $\sim$10-100 pc scales, and (iii) The young Galactic Centre disc(s). We account for multiple encounters between binaries and the MBH, using N-body simulations to model the approach of disc binaries to disruption. 

We are unable to reproduce the observed velocity distribution. 
This conclusion holds for different loss cone refilling mechanisms, including the disc instability, and 
suggests that observed hypervelocity star candidates are not predominantly coming from the Galactic Centre. Alternatively, there may be observational biases against observing the fastest stars. 
In principle,  unique star formation histories or delays in HVS ejections due to relaxation processes may play a role, and explain the HVS high velocity tail deficiency, however, we find that such delays are tightly constrained by the observed flight times of hypervelocity stars.

Binary disruptions by the MBH will leave the companion star of an ejected HVS on a close bound orbit around the MBH \cite{hills1988}. The Galactic Centre S-stars were suggested to originate from such processes \citep{gou+03}. Consequently, the distribution of the orbits of the S-stars should generally be correlated with the distribution of HVS properties. We find that this is not the case, further challenging the Galactic Centre binary disruptions origins for most of the observed HVSs.       

In addition to disruptions, close encounters between binaries and the MBH can produce stellar mergers, especially among the closest binaries \citep{ginsburg&loeb2006}. In fact, multiple close encounters can substantially increase the number of mergers, as described in \citet{bradnick+2017}. Dusty gas clouds like G2 in the Galactic Centre \citep{gillessen+2012,gillessen+2013,ciurlo+2020} could be produced through stellar mergers \citep{prodan+2015}.

The remainder of this paper is organized as follows. In \S~\ref{sec:comp} we describe different sources for binary disruptions, and compare expected velocity distribution for each source to the observations. In \S~\ref{sec:disc}, we discuss additional predictions and physical effects. Finally, we summarize our results in \S~\ref{sec:summ}.

\section{Velocity distribution}
\label{sec:comp}
\subsection{Analytic theory}
\label{sec:analytic}
Binary stars that approach an MBH too closely will be torn apart. The characteristic distance where this occurs is 

\begin{equation}
    r_t=\left(\frac{M_{\rm bh}}{m_{\rm bin}}\right)^{1/3} a_{\rm bin},
\end{equation}
where $M_{\rm bh}$ is the mass of the MBH, $m_{\rm bin}$ is the mass of the binary, and $a_{\rm bin}$ is its semimajor axis. We refer to this distance as the `nominal tidal radius.' In fact, binary disruption is a probabilistic process that depends on the internal orbital phase, such that the disruption probability is not one at $r_t$. Furthermore, binaries may be disrupted a factor of a few beyond the nominal tidal radius \citep{hills1991,bromley+2006,sari+2010}.

Binaries wander in angular momentum space, and some will enter a loss cone of orbits, where the pericentre is smaller than the tidal radius. If the change in a binary's angular momentum per orbit is large compared to the size of the loss cone, it is in the full loss cone regime. Otherwise, it is in the empty loss cone regime. In the latter case, the binary will spend many orbits near the tidal radius prior to being disrupted \citep{merritt2010}. 

In either case, once the binary is tidally disrupted, it is split into two stars. For a parabolic centre-of-mass orbit, one star is bound to the supermassive black hole, while the other is ejected from the Galactic Centre. The typical ejection velocity is 

\begin{equation}
    v_{\rm ej}\approx \left(\frac{M_{\rm bh}}{m_{\rm bin}} \right)^{1/6} \left(\frac{G m_{\rm bin}}{a_{\rm bin}}\right)^{1/2}.
    \label{eq:vej}
\end{equation}
The velocity distribution depends on the semimajor axis distribution of disrupting binaries.
In the empty loss cone regime the semimajor axis distribution of disrupting binaries will correspond to the intrinsic semimajor axis distribution of the population (modulo a logarithmic correction). In the full loss-cone regime, the disruption rate will be proportional to the tidal radius \citep{perets&gualandris2010}. Consequently, for a log-uniform binary semimajor axis distribution,
the velocity distribution of ejected stars is $\propto v_{\rm ej}^{-3}$ ($\propto v_{\rm ej}^{-1}$) in the full (empty) loss cone regime  \citep{rossi+2014}. However the latter scaling does not account for the logarithmic dependence of the disruption rate on the tidal radius. When this is included the distribution in the empty loss cone regime is closer to $v^{-1.2}$.

\subsection{Observational data}
Recently \citet{warren_brown+2018} used Gaia proper motion data to constrain the origin of B stars in the HVS survey. They identified seven stars that were probably ejected from the Galactic Centre, as well as seven stars that are likely runaways from Galactic disk, along with a handful of halo stars. The remaining 21 stars in their sample have ambiguous origin.\footnote{Note that while the \citet{warren_brown+2018} stars are all radial velocity outliers approximately half are still bound to the Galaxy, and not hypervelocity stars.} Here, we compare the velocity distribution from binary disruption in various regimes (e.g. full and empty loss cone) to the velocity distribution of Galactic Centre and ambiguous stars from \citet{warren_brown+2018}.  

The S-star cluster in the Galactic Centre provides additional constraints on the origins of this sample.

\begin{figure}
    \includegraphics[width=\columnwidth]{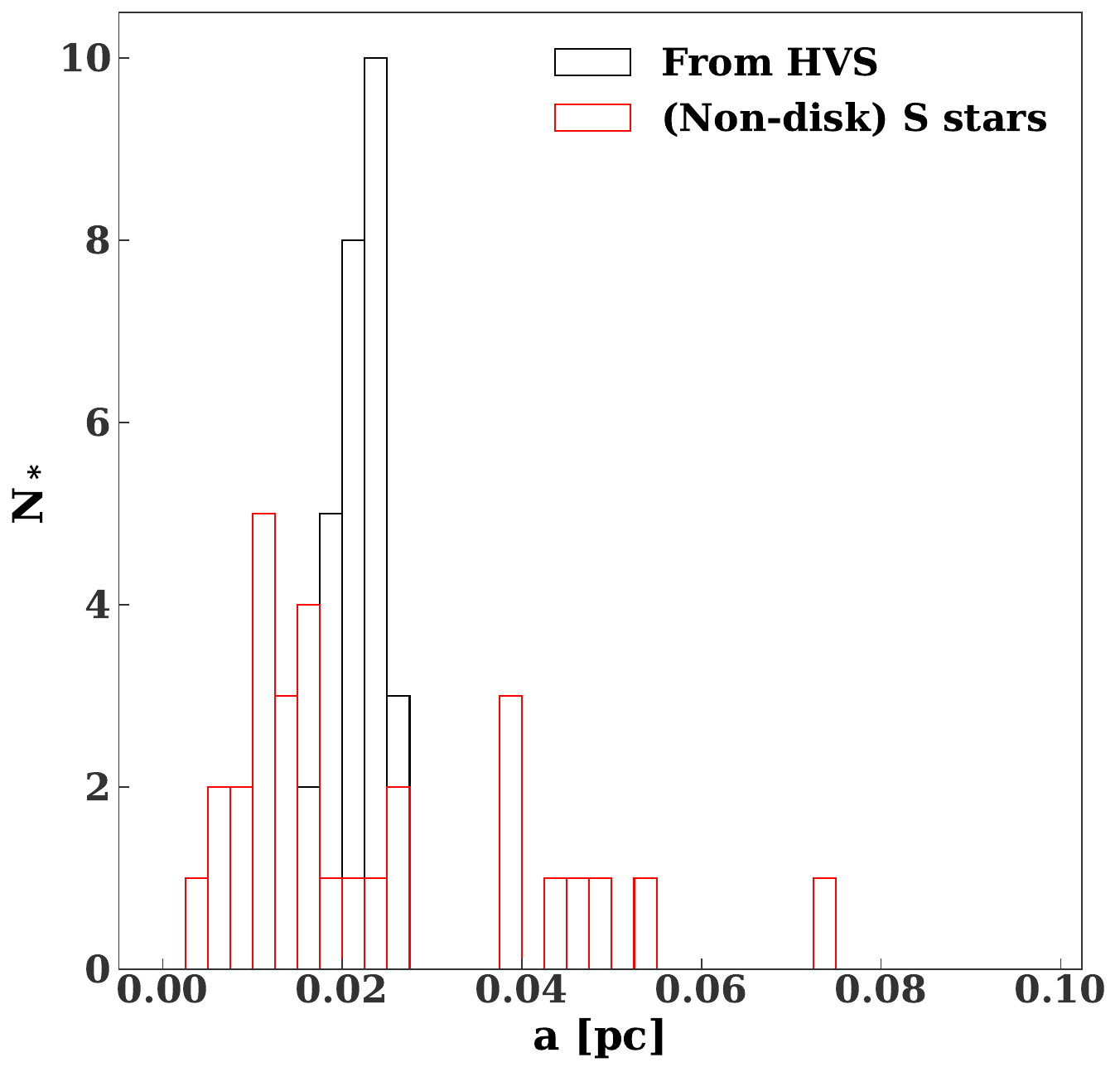}
    \caption{Semimajor axis distribution of hypothetical companion stars for the Galactic Centre and ambiguous stars from \citet{warren_brown+2018}, compared to the observed distribution semimimajor axis of the (non-disk) S-stars from \citet{gillessen+2017}.
    \label{fig:sstarcomp}}
\end{figure}

If the stars were really ejected via binary disruption in the Galactic Centre, their companions should be deposited near this cluster.
\cite{wenbinlu+2020} studied the fastest S5-HVS1 and its former companion; here we discuss the overall sample of HVSs. 
 For each star in the `Galactic Centre' and `ambiguous' samples of \citet{warren_brown+2018}, we use the the Galactic potential described in Appendix~\ref{app:decel} to infer its ejection velocity, and then use this velocity to infer the semimajor axis of the former companion, assuming it is of comparable mass to the ejected star and approximating the centre-of-mass orbit of the progenitor binary as parabolic.
  We also assume the companion has not significantly shifted in semimajor axis. Figure~\ref{fig:sstarcomp} shows a comparison of the inferred semimajor axes and the semimajor axes of the S-stars. The two distributions are inconsistent at high significance,\footnote{For example, a two-sample Kolomgorov--Smirnov test with the S-stars inside 0.03 pc returns a p-value of $3\times 10^{-7}$} suggesting that these populations have different origins. Lower mass companions could in principle alleviate the tension. However, the S-star are B-type stars like the HVS candidates in \citet{warren_brown+2018}, and in fact have comparable or even somewhat greater masses exacerbating the tension \citep{habibi+2017,cai+2018}. The counterparts would have to be dimmer, low-mass stars at smaller semimajor axis that are not presently observed.

\subsection{Numerical study}
We consider a few sources for binary disruptions, spanning different regimes for refilling the binary loss cone. In all cases, we are unable to reproduce the observed velocity distribution from \citet{warren_brown+2018}.

We consider: (i) binaries from a  spherical population in the Galactic Centre and (ii) binaries from a young stellar disk in the Galactic Centre, similar to the one presently observed \citep{levin&beloborodov03, paumard+2006}. For the spherical case, we consider binaries from small (central few pc) and large radii ($\sim 10-100$ pc), where binaries can be scattered to disruption by molecular clouds or larger structures \citep{perets+2007,hamers+17}.
We refer to these cases as the `inner spherical model' and `outer spherical model.' We summarize all of our models in Table~\ref{tab:models}.

\begin{table*}
     \caption{Summary of the models used in this study.}
    \begin{tabular}{l|c|c|c}
        Model & Pericentre distribution & Galactocentric distance & Primary mass function\\ 
        \hline
        \hline
        Inner spherical & From random walk & $\leq 3$ pc & $m^{-2.3}$\\ 
        Outer spherical & Uniform          & $\gsim 10$ pc & $m^{-2.3}$\\
        Disc            & From N-body simulations & $< 0.5$ pc & $m^{-1.7}$\\
        \hline
    \end{tabular}
    \label{tab:models}
\end{table*}

For each source, we construct a mock binary population. Then for each binary in the mock population, we simulate three-body encounters with the supermassive black hole in the Galactic Centre (\sgra), using the \texttt{FEWBODY} code \citep{fregeau+2004}. For the inner spherical and disc models, we explicitly model the binaries' approach to disruption. Repeated encounters can excite the binaries' eccentricities, and catalyze collisions between stars \citep{bradnick+2017}, especially among the closest binaries. Additionally, they may cause binaries to expand in semimajor axis.

We assume stars are ejected continuously and isotropically from the Galactic Centre, and propagate them through a model Galactic potential, as described in Appendix~\ref{app:decel}. If the total age of the ejected star after travelling through the Galactic potential exceeds its main sequence lifetime, it is excluded from further analysis.  We discuss the effects of the star formation history on our predictions in \S~\ref{sec:sfh}

\subsubsection{Inner spherical model}
\label{sec:sphere}
We sample binary properties from the following distributions:

\begin{enumerate}
\item The primary mass is drawn from an $m^{-2.3}$ distribution between 1 and 10 $M_{\odot}$.
\item The binary semimajor axis, eccentricity, and mass ratio are drawn from the distributions in Section 9 of \citet{moe+2017}. These distributions are based on fits to binary properties from many different surveys, and account for observational selection effects. The semimajor axis distribution is close to log-uniform for B-type binaries, while the eccentricity distribution goes from a nearly thermal distribution at large periods to circular at small periods.
\item The inclination is drawn from an isotropic distribution. 
\end{enumerate}
In order to reach a Galactocentric distance of $20$ kpc with a velocity of 275 km s$^{-1}$, a star must be ejected from the Galactic Centre with a velocity $\gsim$750 km s$^{-1}$. Typically, only binaries with semimajor axes less than $\sim$1 au produce such fast stars (see equation~\ref{eq:vej}). However, in rare cases wider binaries in the full loss cone regime can as well (such large energies are accessible to binaries with pericentres that are significantly smaller than the nominal tidal radius; see Figure 7 in \citealt{sari+2010}). To account for this effect we use a maximum semimajor axis of 15 au. 

As the observed sample in \citet{warren_brown+2018} are B stars with a few solar masses, we only include $3-4 M_{\odot}$ in our observational comparisons.

The binary population is assumed to have a relaxed $r^{-1.75}$ density profile \citep{bahcall&wolf1976}. For each binary we sample a Galactocentric radius (between 0.01 and 3 pc) from this distribution. We also ran models with an $r^{-0.5}$ core (with the same relaxation time at the outer boundary). The flatter profile had a minor effect on the velocity distribution in the halo.

We then simulate a random walk in angular momentum space for each binary. The age of the binary at the start of the random walk is a random number between 0 and the main sequence lifetime of the primary star.  The initial angular momentum is drawn from a steady-state distribution, $f(J)\propto J \ln (J/J_{\rm lc})$ \citep{merritt2010}.\footnote{This is the steady-state distribution for the empty loss cone regime. Here $J_{\rm lc}$ is the angular momentum of an orbit that grazes the nominal tidal radius, $r_t$.} In order to capture the binary's approach to disruption, we reject pericentres that are less than five times the nominal tidal radius, $r_t$. Also, if the diffusion timescale of a binary exceeds the remaining main-sequence lifetime of its primary, we proceed to the next one. Here, the diffusion timescale is 

\begin{equation}
    t_{\rm diff}=\frac{(J-J_{\rm o})^2}{J_c^2} t_{\rm rx},
\end{equation}
where $J$ is the binary's angular momentum, $J_c$ is the circular angular momentum, $J_{\rm o}$ is the angular momentum corresponding to a pericentre of $5 r_t$, and $t_{\rm rx}$ is the relaxation time. For a Bahcall--Wolf cusp,

\begin{equation}
    t_{\rm rx}=t_{\rm rx,pc} \left (\frac{r}{\rm pc}\right)^{0.25}.
\end{equation}
The normalization is a free parameter.

After time $\delta t$, the angular momentum changes by 
\begin{equation}
    \delta J=\xi \left(\frac{\delta t}{t_{\rm rx}}\right)^{1/2} J_c+ \frac{J_c^2 \delta t}{2 t_{\rm rx} J}, 
\end{equation}
where $\xi$ is drawn from a normal distribution with unit variance \citep{risken1989}. The step-size is the minimum of the orbital period and 
$t_{\rm diff}/20$. Similar approaches have been used extensively in the literature to study angular momentum diffusion (see e.g. \citealt{hopman&alexander2005, bar-or&alexander2016,wenbinlu+2020,tep+2021}).

For each random walk, all pericentres within five times the (initial) nominal tidal radius are recorded.
If the minimum pericentre is greater than three times the nominal tidal radius, then the binary has a small chance of being disrupted 
\citep{hills1991,bromley+2006,sari+2010, generozov&madigan2020}, and we proceed to the next one.
Otherwise, we simulate all close encounters with the \texttt{FEWBODY} code, stopping if the binary is disrupted. 

In each encounter, the binary starts a factor of fifty from its (current) nominal tidal radius. (Binaries with apocentre smaller than fifty times the tidal radius are rejected). The total integration time is three times the time necessary for the binary to reach the pericentre of its orbit. 
The orbital elements of the binary are then propagated unchanged to the next close encounter, except for the mean anomaly, which is randomized.

\begin{figure*}
    \includegraphics[width=0.66\columnwidth]{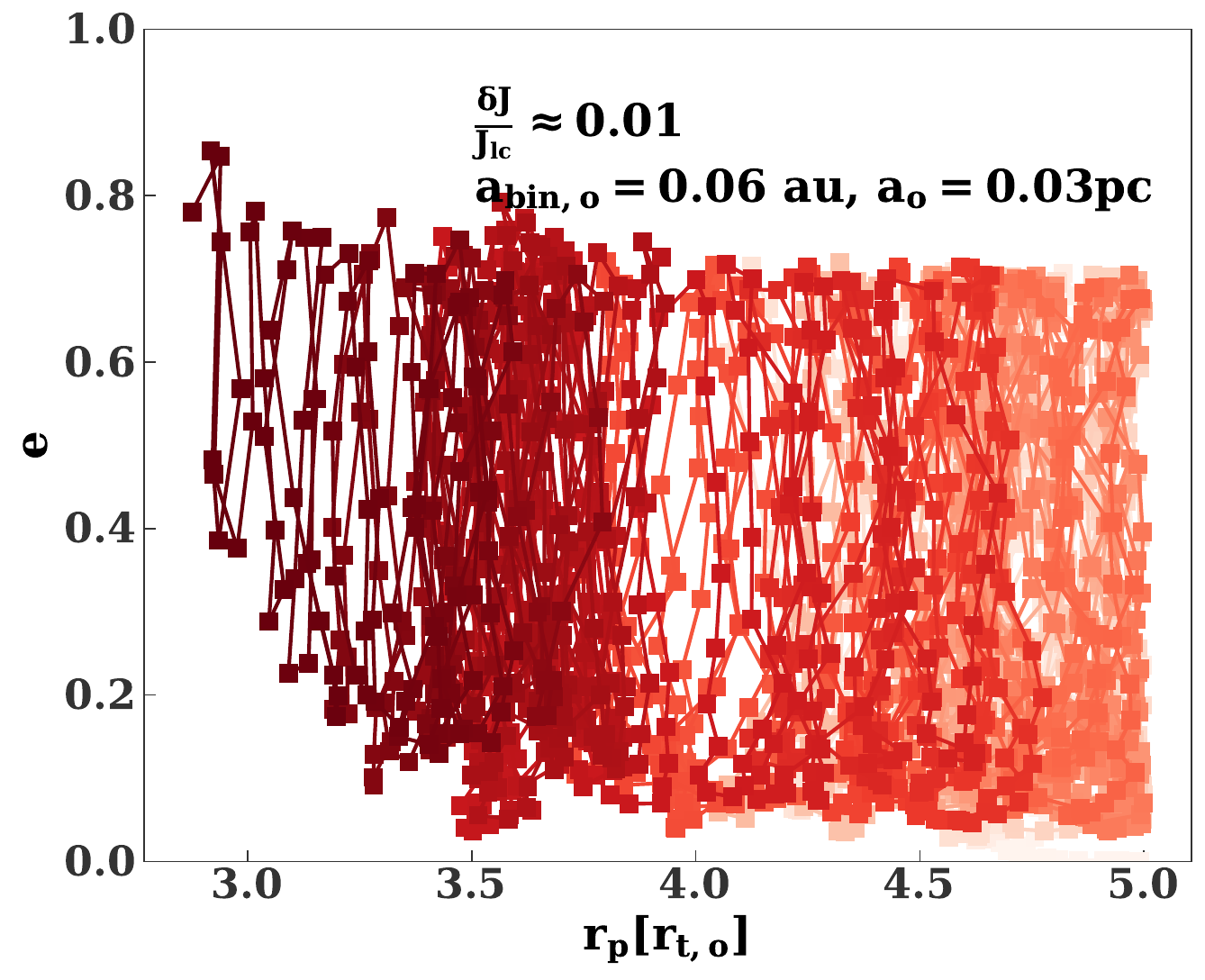}
    \includegraphics[width=0.66\columnwidth]{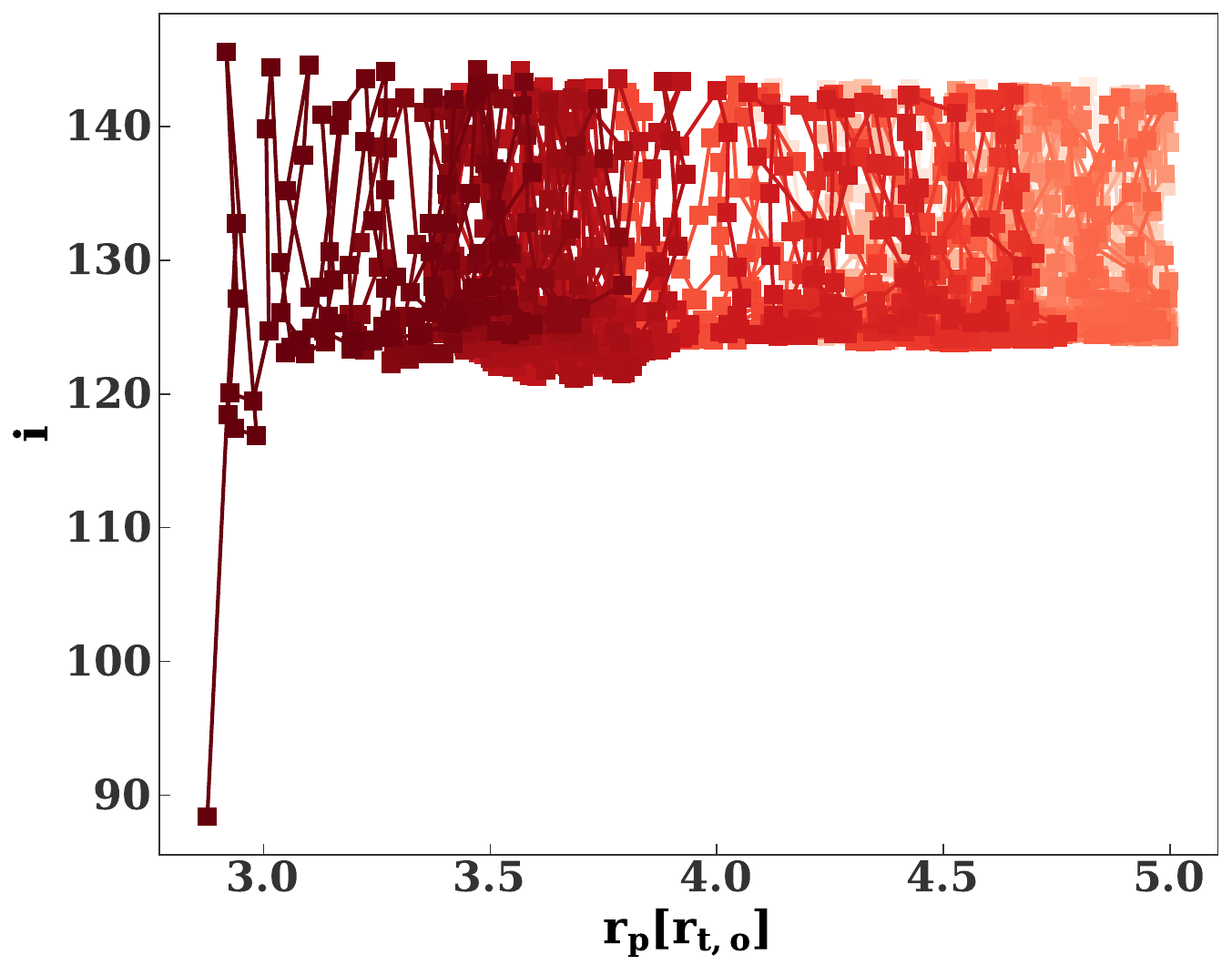}
    \includegraphics[width=0.66\columnwidth]{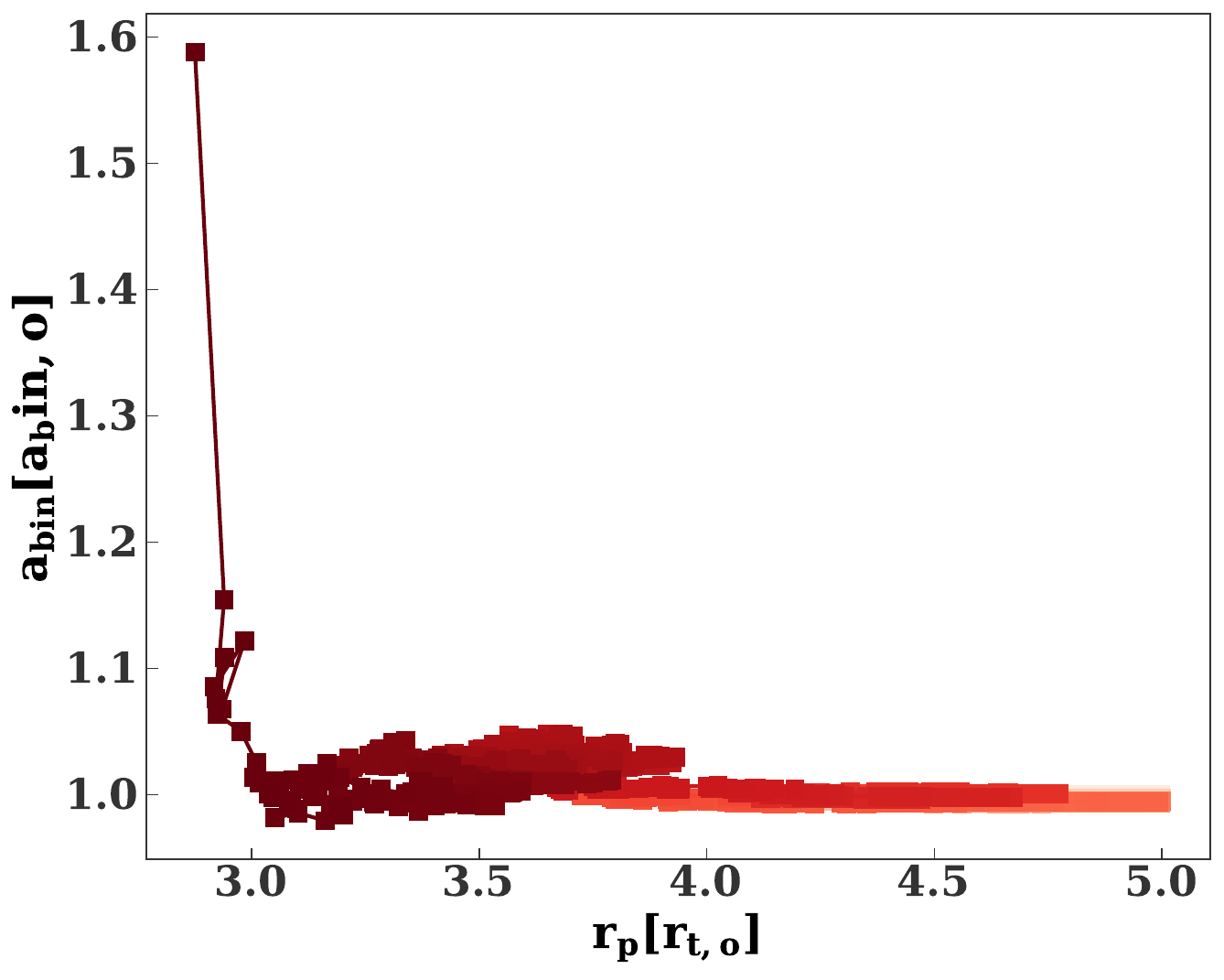}
    \includegraphics[width=0.66\columnwidth]{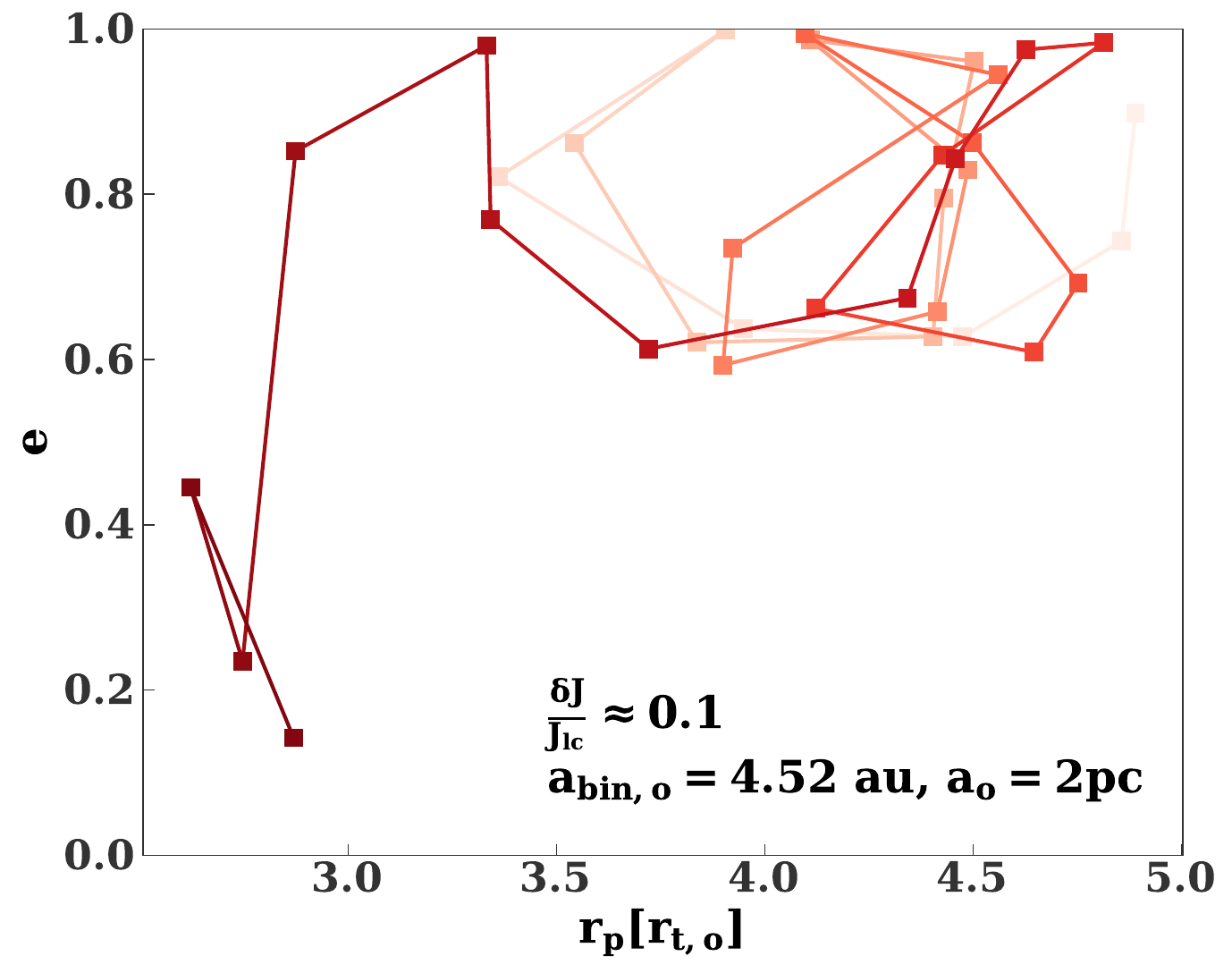}
    \includegraphics[width=0.66\columnwidth]{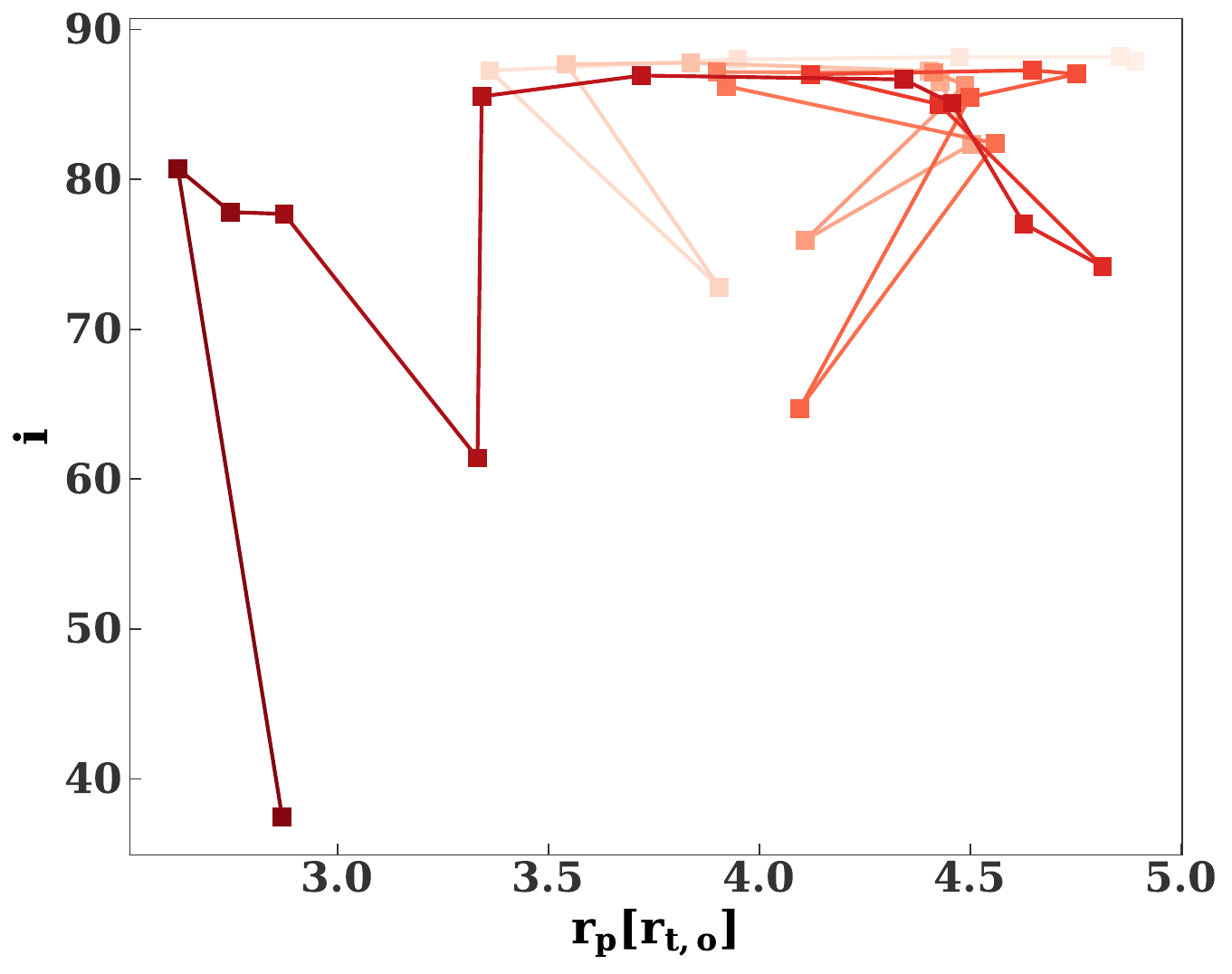}
    \includegraphics[width=0.66\columnwidth]{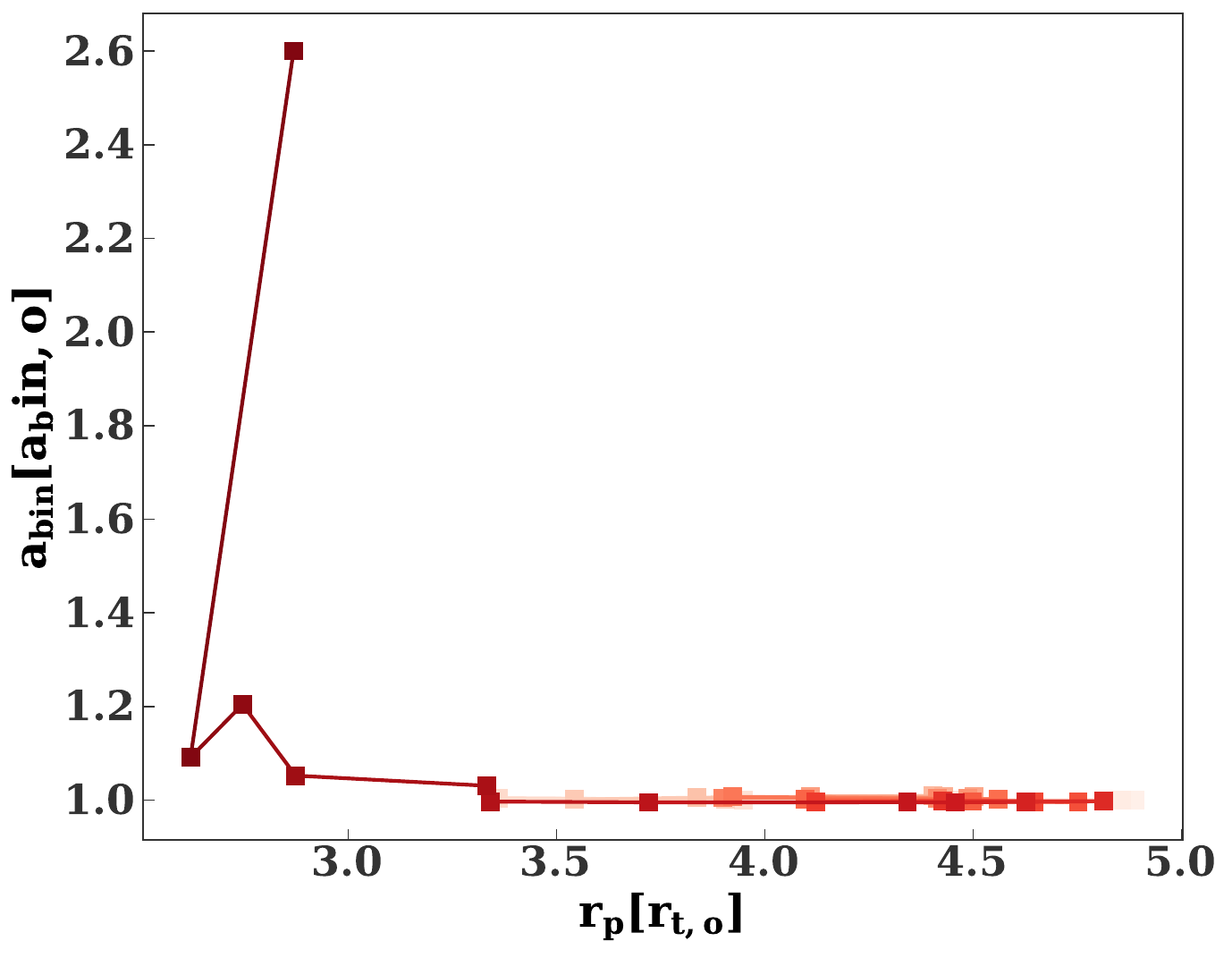}
    \includegraphics[width=0.66\columnwidth]{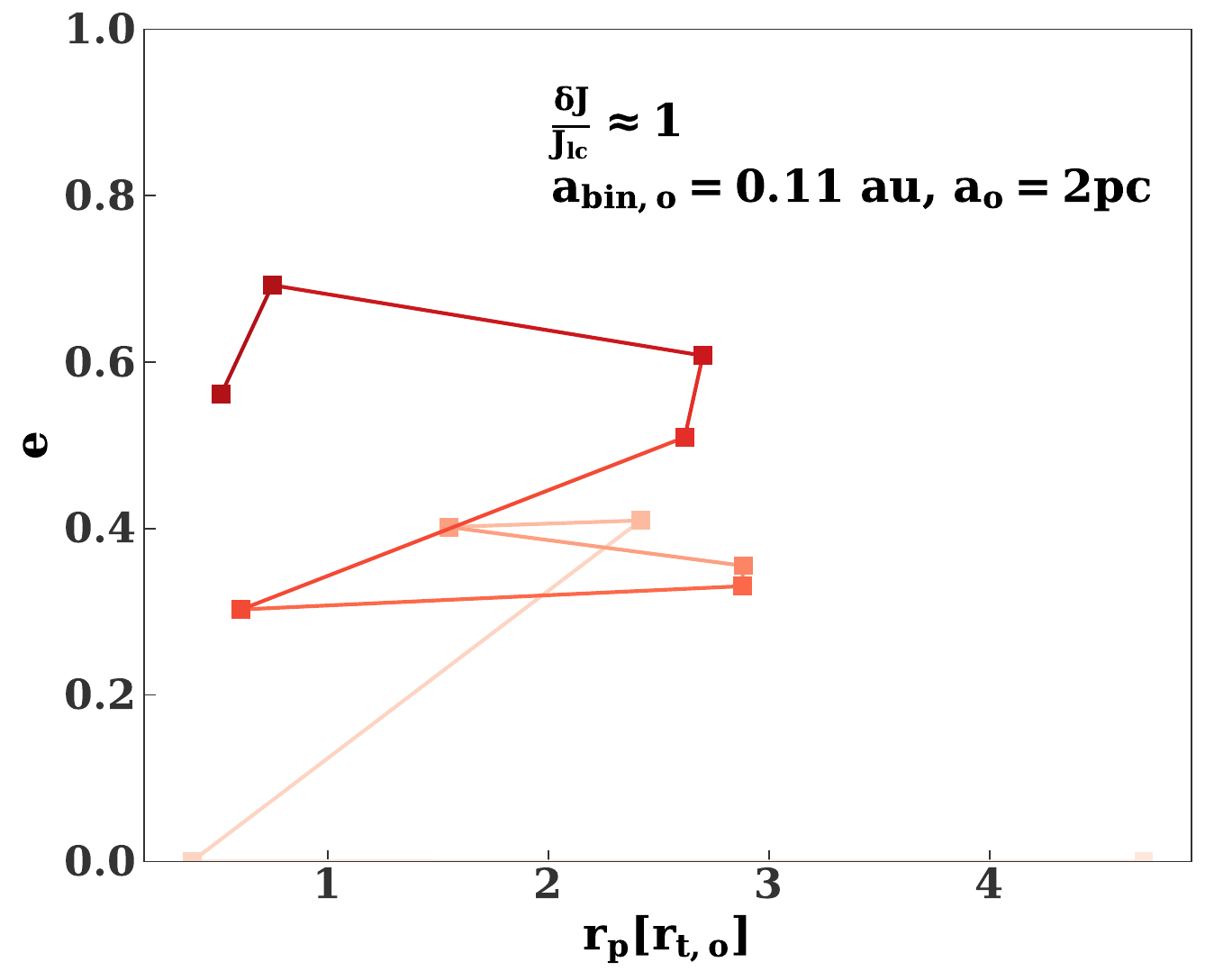}
    \includegraphics[width=0.66\columnwidth]{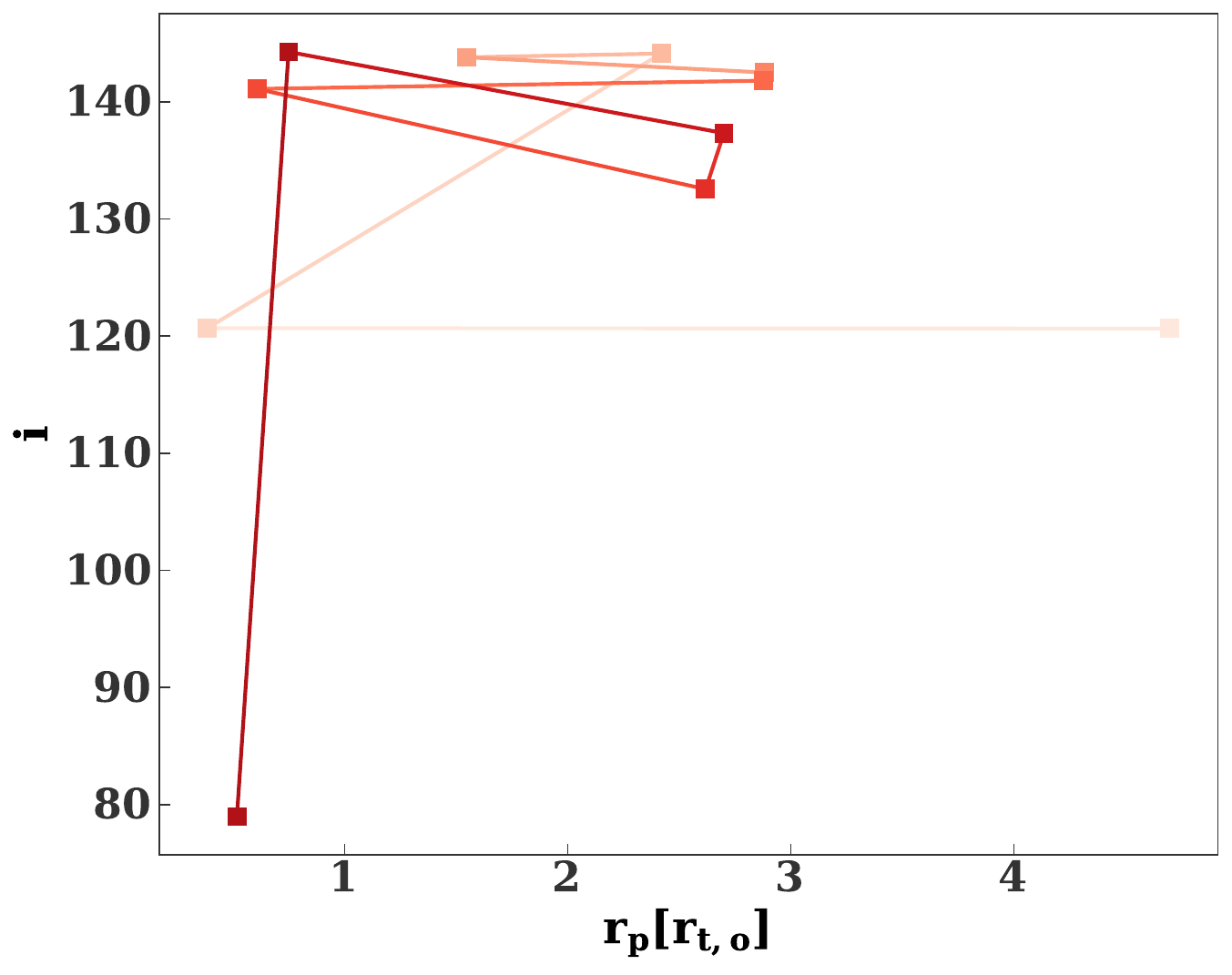}
    \includegraphics[width=0.66\columnwidth]{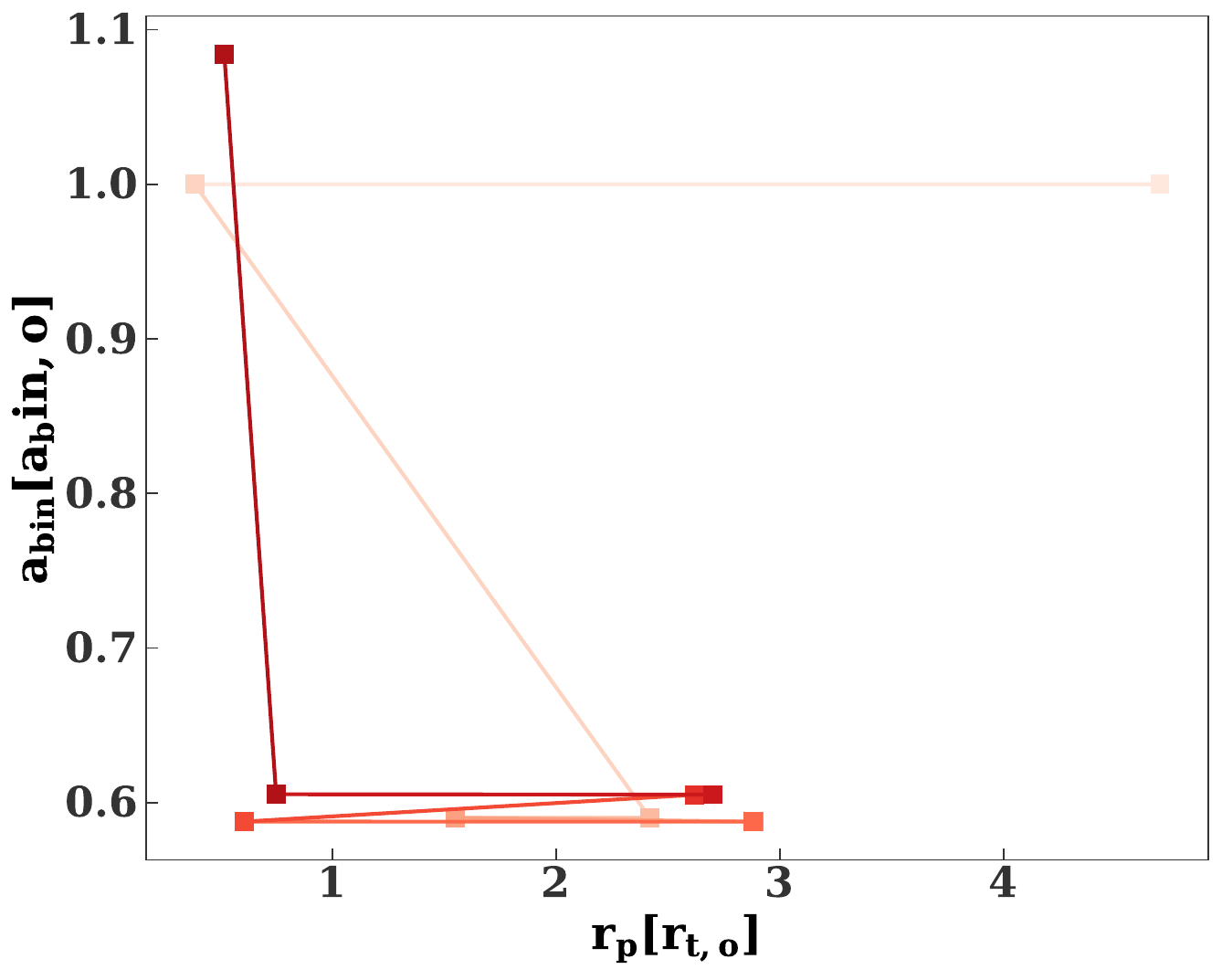}
    \caption{Evolution of binary eccentricity (\emph{left panels}), inclination (\emph{middle panels}) and semimajor axis (normalized to the initial semimajor axis; \emph{right panels}) from three binary disruption simulations. The horizontal axis shows the pericentre in units of the initial nominal tidal radius. Each point corresponds to a close encounter, with darker colors corresponding to later times. The top panel shows a binary deep in the empty loss cone regime, where many close encounters occurring before disruption. The subsequent panels correspond to binaries, closer to the full loss cone regime.}
    \label{fig:examples}
\end{figure*}

Figure~\ref{fig:examples} shows examples of binary evolution from our simulations. In the top panel, the binary is deep in the empty-loss cone regime, and takes many steps in angular momentum, prior to disruption. In the process, it experiences ZLK-like (von Zeipel, Lidov, Kozai; \citealt{lidov1962, kozai1962}; see \citealt{naoz2016} for a review) oscillations in eccentricity and inclination, as suggested to occur by \cite{antonini&perets2012}. Binaries closer to the full loss cone regime (middle and bottom panels), that undergo fewer encounters prior to disruption, and the eccentricity evolution is more chaotic. The semimajor axis of the binary can expand by a factor of a few prior to disruption, as shown in the right panels. 

\begin{figure}
    \includegraphics[width=\columnwidth]{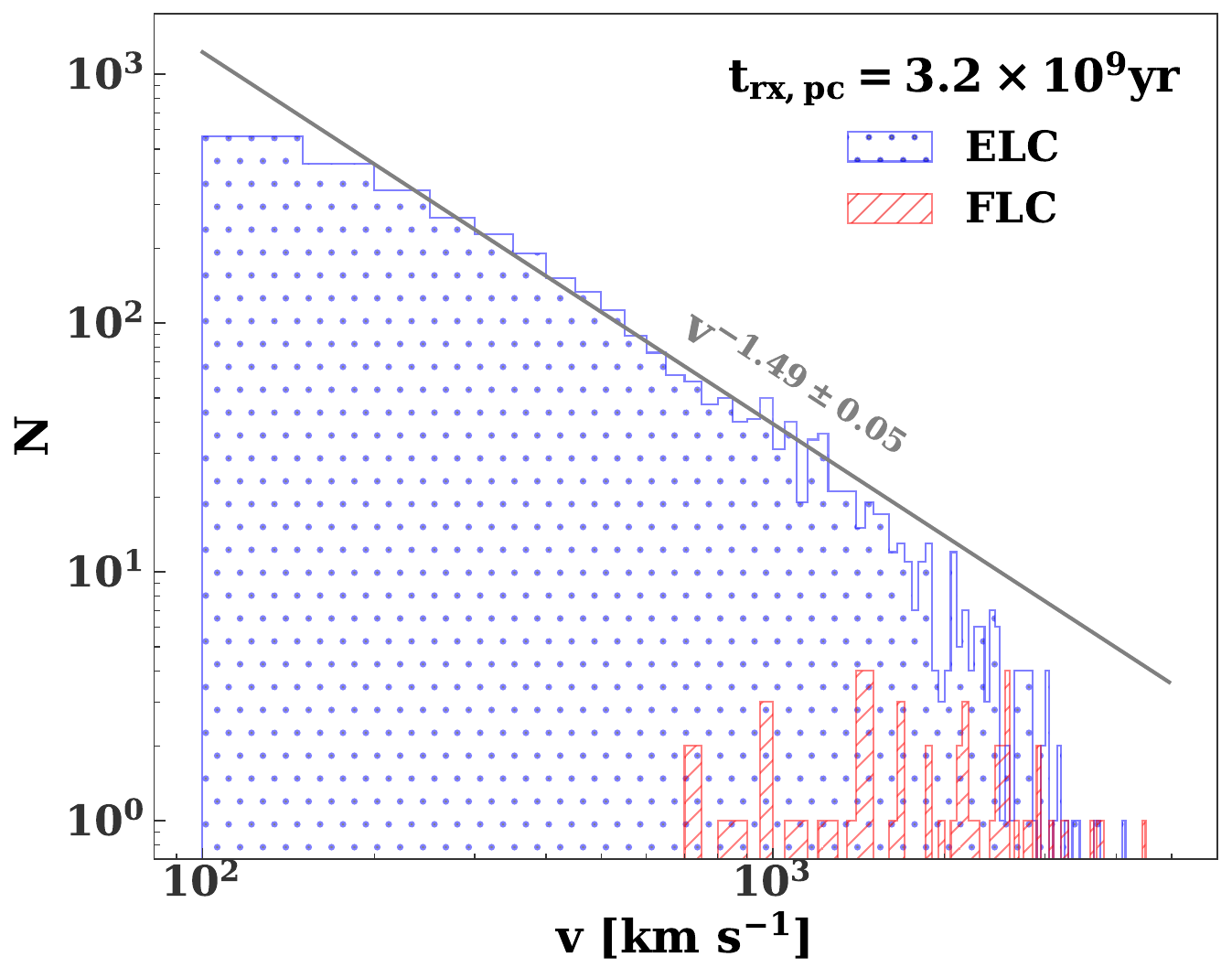}
    \includegraphics[width=\columnwidth]{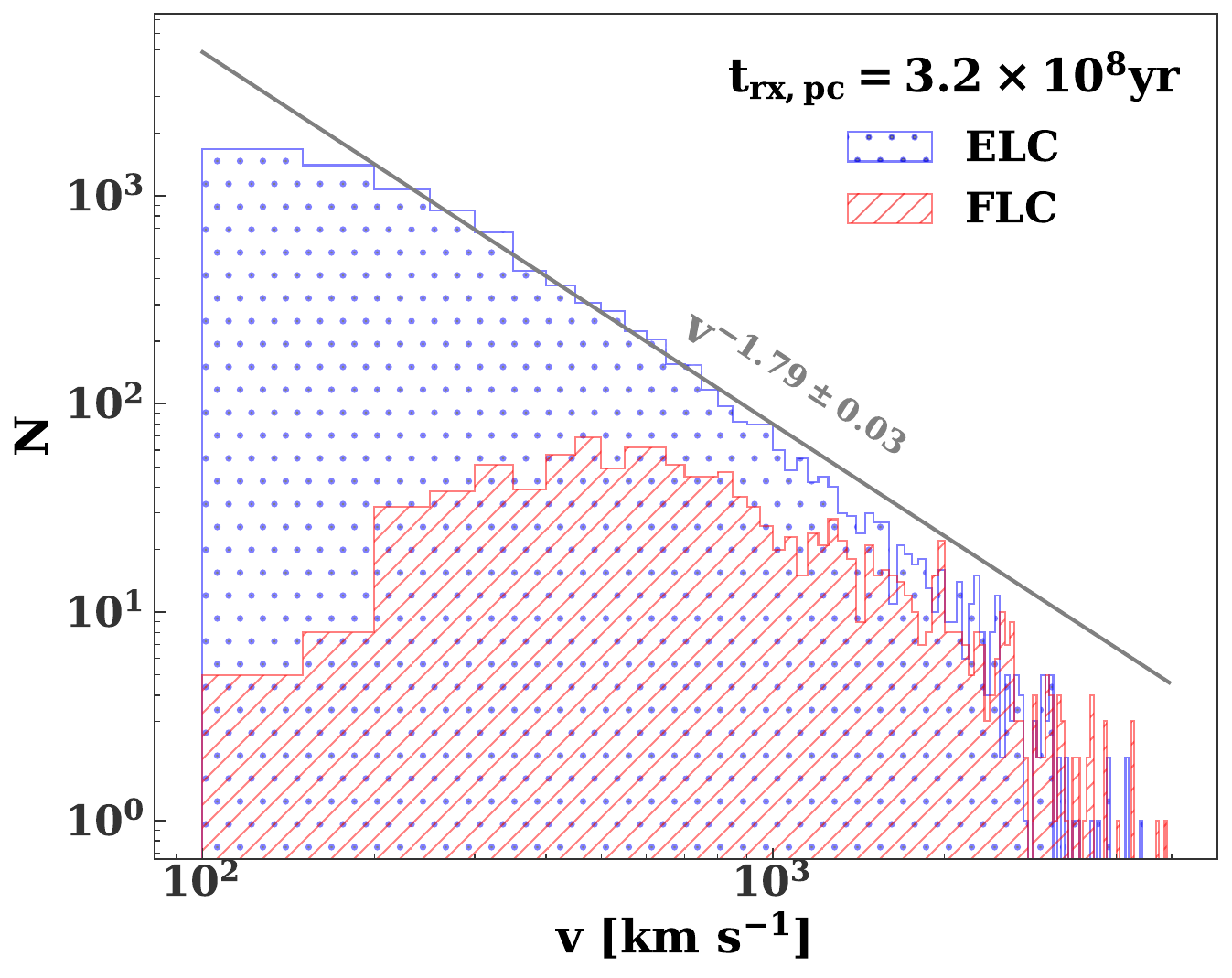}
    \caption{Velocity distribution for ejected stars in the inner spherical model. The two panels correspond to different relaxation times. The red, hatched (blue, dotted) histogram shows the contribution of binaries in the full (empty) loss cone regime. The gray lines correspond to the best fit power-law to the velocity distribution between 200 and 1000 km s$^{-1}$.
    \label{fig:pdf}}
\end{figure}

Figure~\ref{fig:pdf} shows velocity distribution of $3-4 M_{\odot}$ stars after escaping the MBH potential for two different relaxation times in the inner spherical model.
 Our models include binaries coming from different regions, where those coming from the inner regions are closer to the empty loss-cone regime and those further out are closer to the full loss-cone regime. In particular, binaries are in the full loss cone regime if 
 
 \begin{equation}
    \left(\frac{P}{t_{\rm rx}}\right)^{1/2} J_c>J_{\rm lc},
\end{equation}
where $P$ is the orbital period of the binary centre-of-mass, $t_{\rm rx}$ is its relaxation time, and $J_{\rm lc}$ is loss cone angular momentum corresponding to the nominal tidal radius.
 
 As discussed in \S~\ref{sec:analytic} the different dependence of disruption rate on semimajor axis in these regimes leads to different velocity distributions: $v^{-1}$ in the empty loss cone regime and $v^{-3}$ in the full loss cone regime. We therefore expect a distribution in between these two extremes. Indeed, the power-law index of the velocity distribution is $\approx-1.5$ ($-1.8$) between 200 and 1000 km s$^{-1}$ in the top (bottom) panel of Figure~\ref{fig:pdf},\footnote{At lower velocities the MBH potential may flatten the distribution. At higher velocities the distribution would be affected by the truncation of semimajor axis distribution.}. At closer Galactocentric radii, the loss cone becomes emptier, and the velocity distribution becomes flatter. For binaries in the central parsec, best fit power-law index is $-1.17 \pm 0.08$ for $t_{\rm rx, pc}=3.2\times 10^9$ yr.

Next, we propagate ejected stars through the Galactic potential (following the the procedure in Appendix~\ref{app:decel}). Figure~\ref{fig:vdist} shows the cumulative velocity distribution of $3-4 M_{\odot}$ stars reaching Galactocentric distances between 20 and 140 kpc, while on the main sequence. 
The observed distribution from \citet{warren_brown+2018} (\emph{thin, blue line}) differs significantly (two-sample KS-test p-value of $10^{-14}$) from the theoretical distributions at both relaxation times,  with $\sim 60\%$ of stars above 700 km s$^{-1}$ in the latter.

\begin{figure}
    \includegraphics[width=\columnwidth]{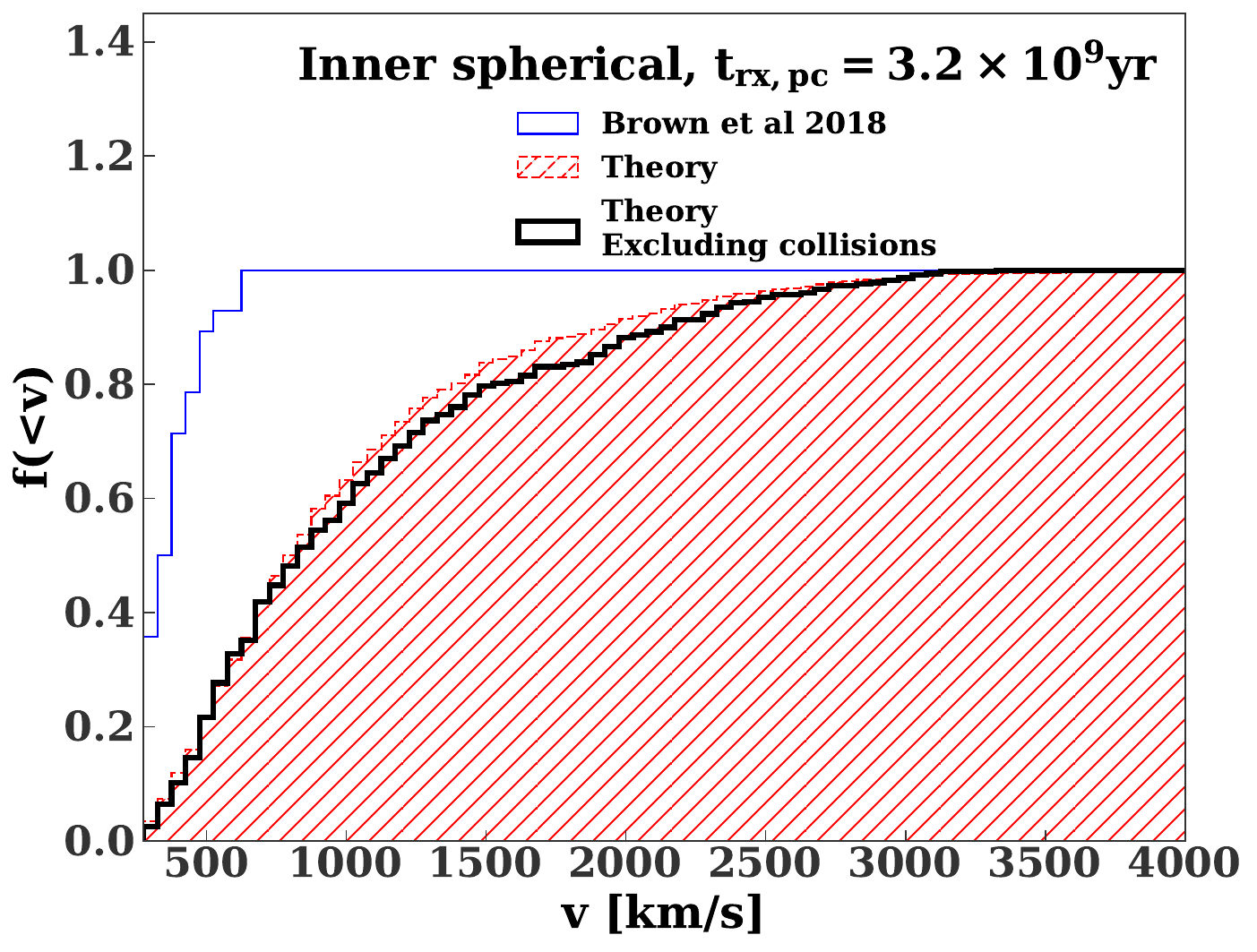}
    \includegraphics[width=\columnwidth]{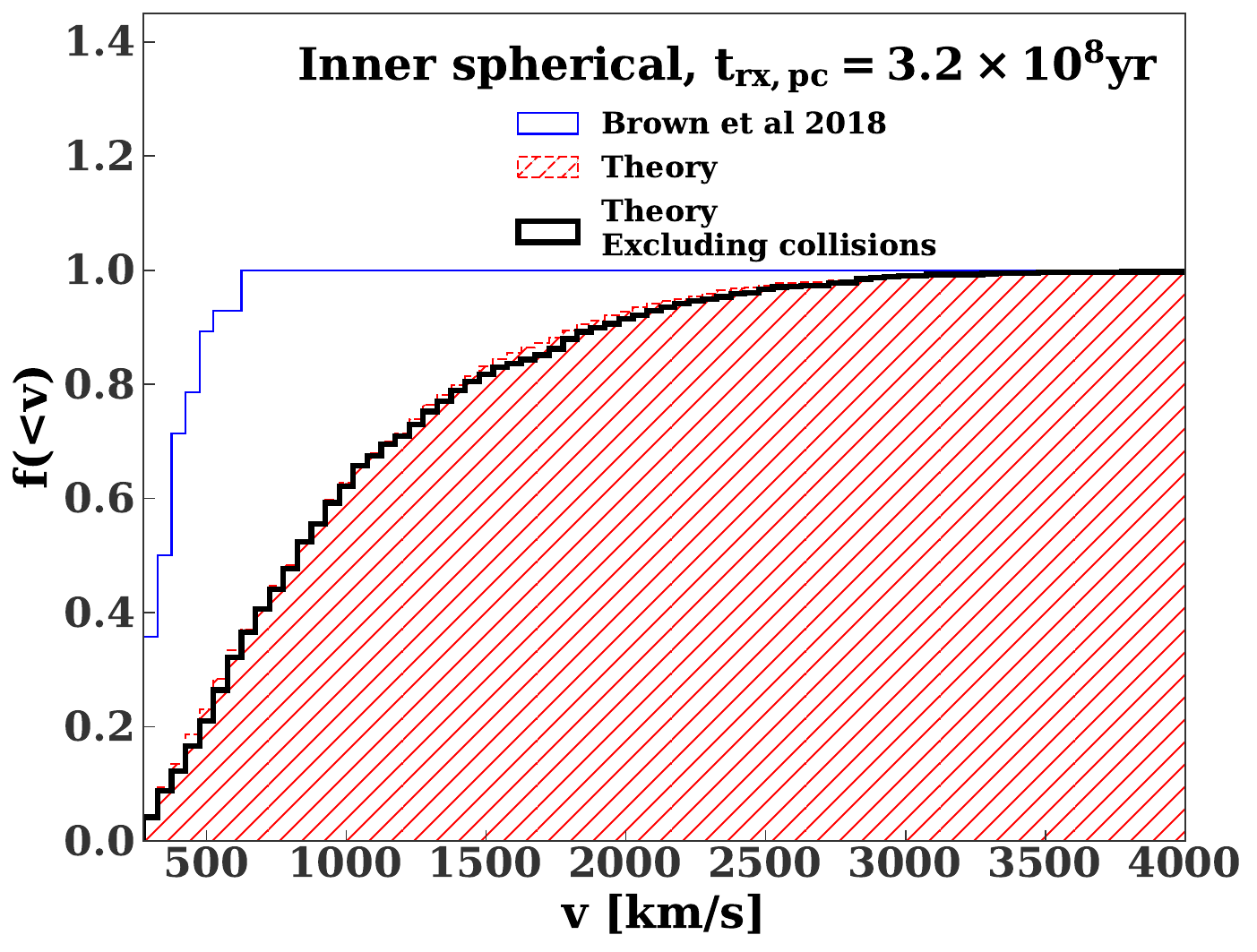}
    \caption{Cumulative velocity distribution of main sequence, $3-4 M_{\odot}$ stars from our inner spherical model after propagation through the Galactic potential, compared to the \citet{warren_brown+2018}  sample (\emph{thin, blue}). To illustrate the effects of collisions we show the velocity distribution with (\emph{thick, black}) and without (\emph{hatched, red}) stellar collisions excluded.
    We show results for two different normalizations of the relaxation time ($t_{\rm rx,pc}=3.2\times 10^9$ yr in the top panel and $t_{\rm rx, pc}=3.2\times 10^8$ yr in the bottom panel). }
    \label{fig:vdist}
\end{figure}

\subsubsection{Outer spherical model}
We now model disruptions of binaries scattered in from large scales by massive perturbers \citep{perets+2007}. 

We use the same distributions for the internal binary properties as in the preceding section, but assume a nearly parabolic centre-of-mass orbit. We assume binaries are in the full loss cone regime, and draw a pericentre from a uniform distribution extending from zero to three times the maximum nominal tidal radius of the whole binary population. If the selected pericentre is more than three times the nominal tidal radius of the binary, we proceed to the next one. Otherwise, we simulate a close encounter. The age of the binary at disruption is uniformly distributed between 0 and the main sequence lifetime of the primary star.

\begin{figure}
    \includegraphics[width=\columnwidth]{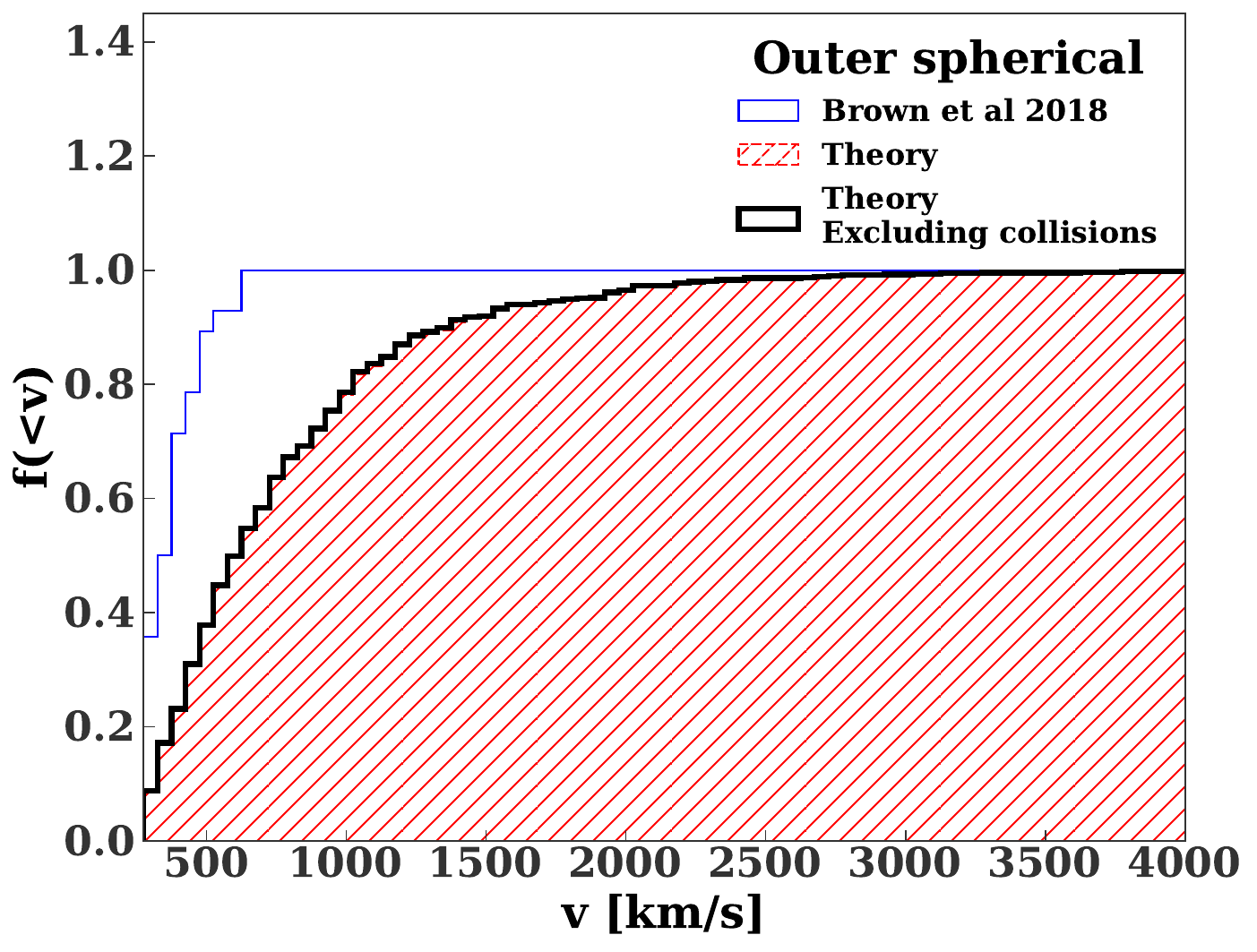}
    \caption{Cumulative velocity distribution from the outer spherical model after propagation through the Galactic potential, compared to the velocity distribution of the \citet{warren_brown+2018} sample. In this case, all binaries are (by construction) in the full loss cone regime.}
    \label{fig:flc}
\end{figure}

Figure~\ref{fig:flc} shows the velocity distribution in the case is also inconsistent with the observations (two-sample KS-test p-value of $\approx10^{-8}$), with $\sim 40\%$ of stars above 700 km s$^{-1}$.

\subsubsection{Binaries from young stellar disc}
    Previous work has shown that a lopsided, eccentric disk in the Galactic Centre undergoes a secular gravitational instability that excites some of its orbits to high eccentricities. Binaries on such orbits can be tidally disrupted \citep{madigan+2009, generozov&madigan2020}.

    Here we use eccentric disk simulations, similar to those in \citet{generozov&madigan2020} to quantify the velocity distribution of ejected stars from such disks. These simulations are initialized with 100 stars on eccentric orbits around a supermassive black hole. The disk extends from $\sim$0.05 to 0.5 pc, and all stars start with an eccentricity of 0.7. The disk starts out apsidally aligned, and then spreads out due to differential precession from a background stellar density profile. Here we assume an $r^{-1.5}$ density profile with $5\times 10^5 M_{\odot}$ within 1 pc.

\begin{figure}
        \includegraphics[width=\columnwidth]{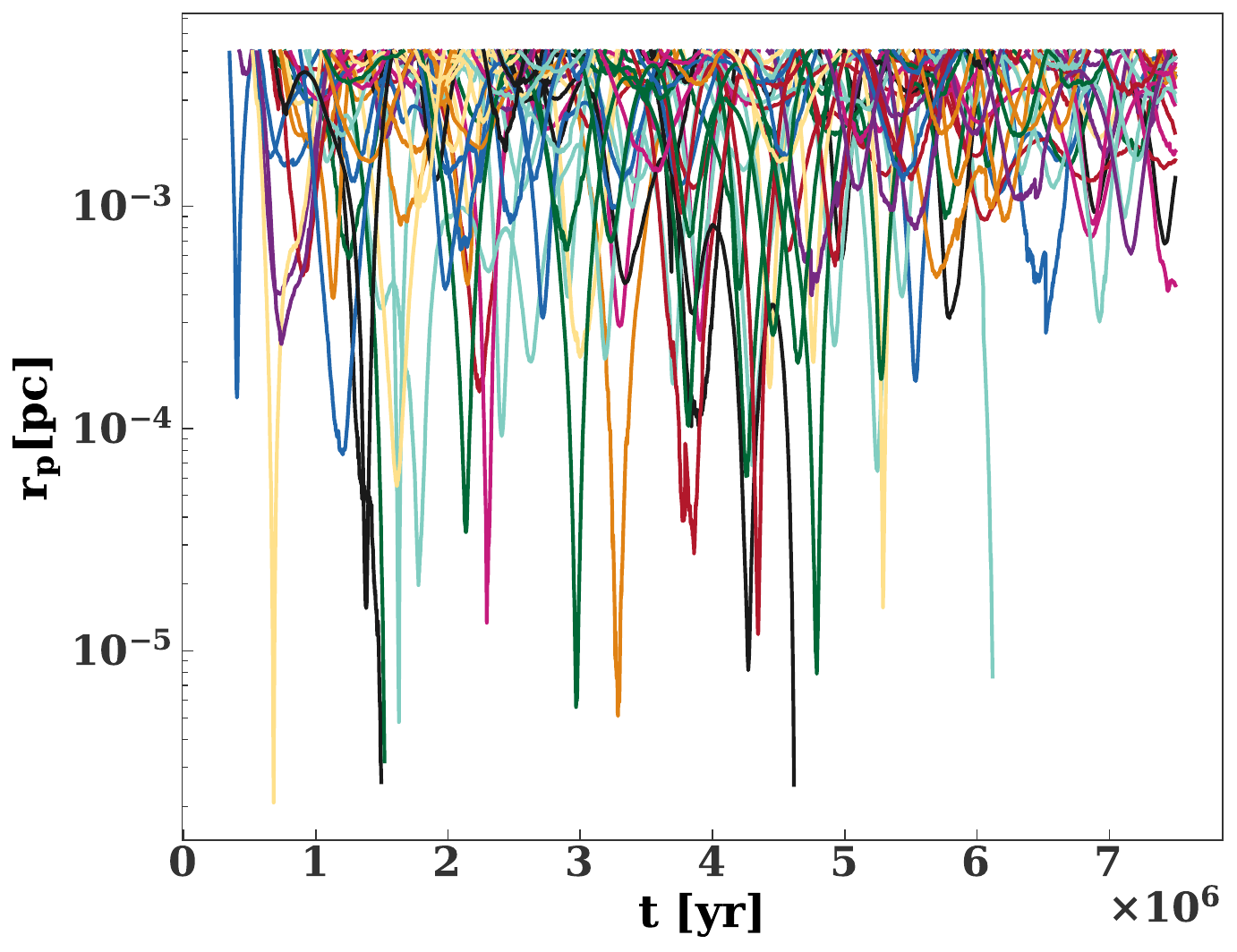}
        \caption{Pericentre versus time for particles that experience close encounters with the central black from eccentric disk simulations (see text for details). Each line corresponds to a different particle.}
        \label{fig:encounters}
    \end{figure}

\begin{figure}
    \includegraphics[width=\columnwidth]{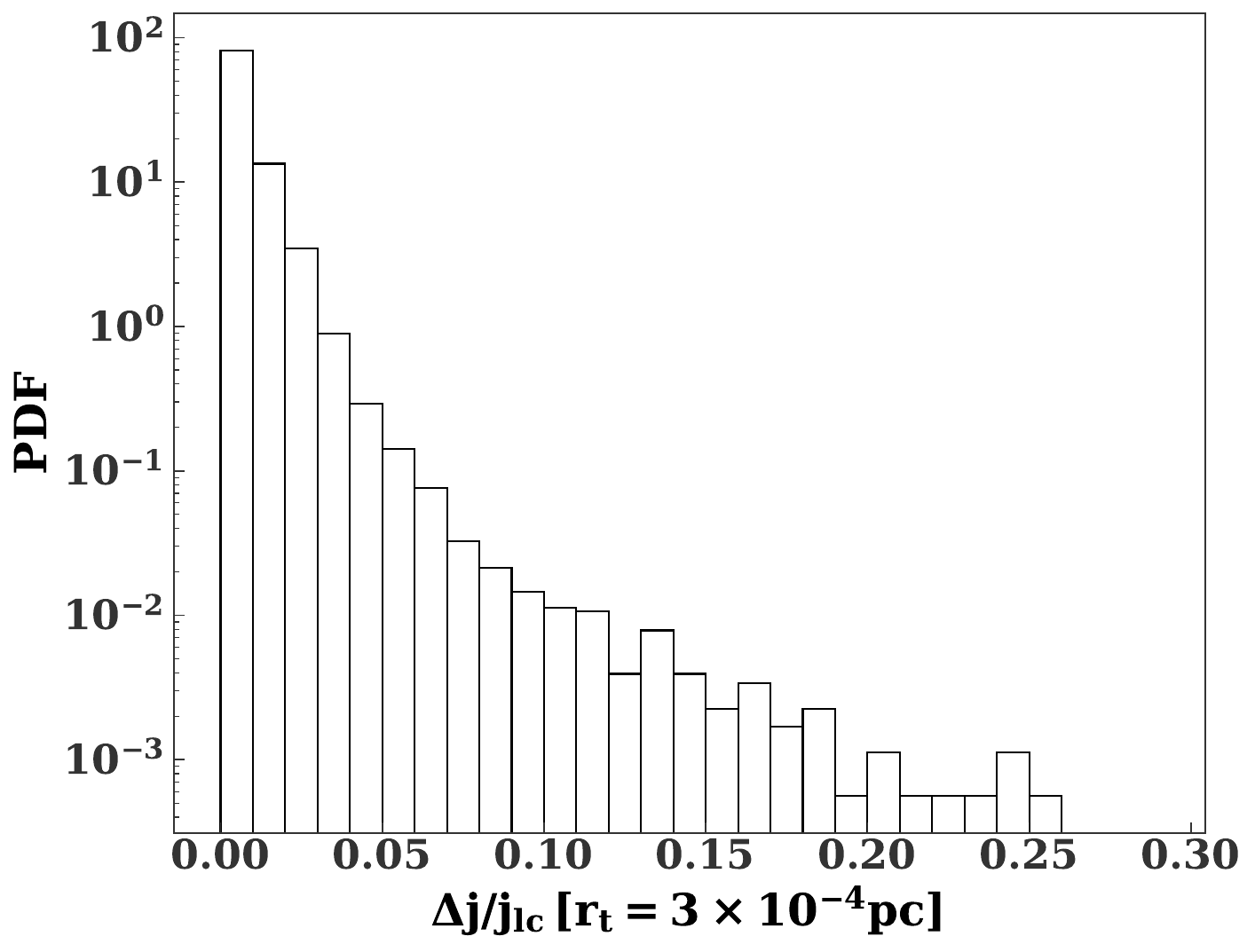}
    \caption{Distribution of $\Delta j$, the change in angular momentum between consecutive encounters in Figure~\ref{fig:encounters}. This is normalized by the loss cone angular momentum for a tidal radius of $3\times 10^{-4}$ pc. }
    \label{fig:deltaJ}
    \end{figure}

    Figure~\ref{fig:encounters} shows all close encounters from these disk simulations.
    Figure~\ref{fig:deltaJ} shows the distribution of $\delta j$, the change in angular momentum between consecutive encounters, normalized by the the loss cone angular momentum ($j_{\rm lc}$) for an assumed tidal radius of $3\times 10^{-4}$ pc.\footnote{This is comparable to the the tidal radius of a $10 M_{\odot}, 1$ au binary.} The size of the angular momentum jumps would place the disk in the empty loss cone regime. However, the angular momentum does not undergo a random walk, as changes are coherent over many orbital periods.
    
We follow a similar procedure to that in \S~\ref{sec:sphere} to model the production hypervelocity stars from the disk. Though, we do make somewhat different assumptions about the binary population. Firstly, we consider an $m^{-1.7}$ mass function for the binaries, motivated by observations of the mass function in the young disk in the Galactic Centre \citep{lu+2013}. Secondly, we consider a binary semimajor axis distribution truncated at 2 au, instead of 15 au. Binaries wider than $\sim 2$ au will only produce hypervelocity stars when the loss cone is full, which is not the case here. Finally, as disruptions occur promptly after disc formation \citep{madigan+2009,generozov&madigan2020}, the stars are on the ZAMS at disruption in this case.

For each binary we select a random set of consecutive, close encounters from Figure~\ref{fig:encounters} with decreasing pericentres. For each encounter,
we have the complete set of orbital elements for the (single) star in the disk simulation.
We replace the single star with a binary, and simulate repeated close encounters.

\begin{figure}
    \includegraphics[width=\columnwidth]{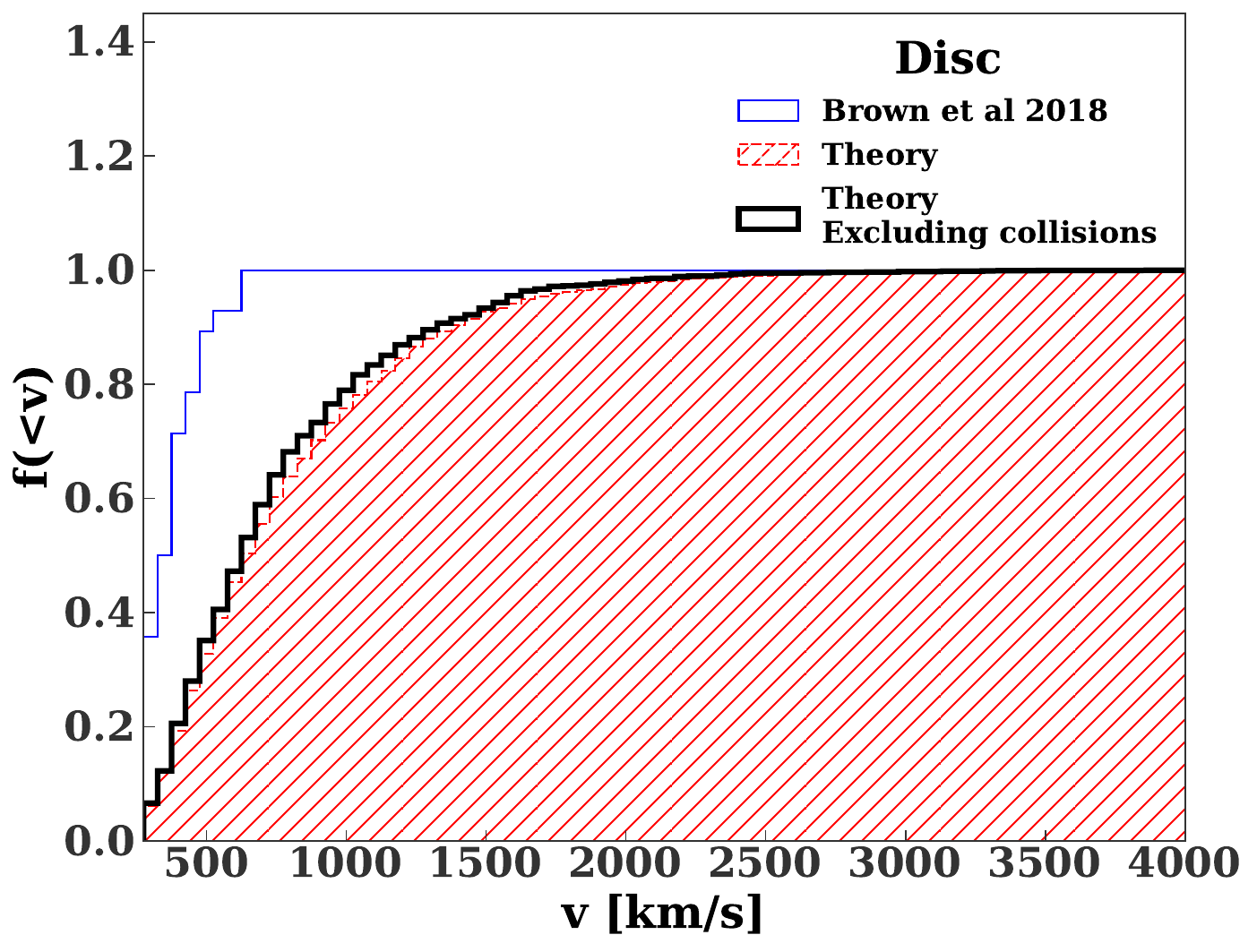}
    \caption{Cumulative velocity distribution for the young disk after propagation through the Galactic potential, compared to the velocity distribution of the \citet{warren_brown+2018} sample (\emph{thin, blue})
    \label{fig:diskv}}
\end{figure}

The velocity distribution of stars ejected by the eccentric disk instability is shown in Figure~\ref{fig:diskv}, along with the velocity distribution of all hypervelocity star candidates and `ambiguous' stars from \citet{warren_brown+2018} in blue. Again, the theoretical distribution does not match the observations (two-sample KS-test p-value $\approx 10^{-9}$), with $\sim 40\%$ of stars above 700 km s$^{-1}$.

\section{Discussion}
\label{sec:disc}
\subsection{Collisions}
Collisions are relatively rare for the outer spherical model, only occurring in $\sim 1\%$ of cases. 
In contrast, $\sim$20-30\% (50\%) of binary stars collide in the inner spherical model (disc). We now discuss such cases.

Collisions can reduce the ejection rate of hypervelocity stars by a factor of a few. However, they do not significantly affect the shape of the velocity distribution as shown in Figures~\ref{fig:vdist}, \ref{fig:flc}, \ref{fig:diskv}, in spite of the higher collision probability of close binaries. This is because stars reaching the halo with velocities above 275 km s$^{-1}$  originate from binaries with a relatively narrow range of semimajor axes: $\sim$0.06 to $\sim$0.6 au. Over this range the collision probability only varies between $\sim$40 and $\sim$60\%. In fact, the effect of collisions on the velocity distribution is completely dominated by noise from the randomized stellar ages. Close encounters where the binary stars narrowly miss each other can also affect the velocity distribution via tidal effects \citep{bradnick+2017}. 
We have performed new binary disruption simulations for our disc model, allowing the binary semimajor axis and eccentricity to evolve according the equilibrium tide model described in \citet{bradnick+2017} (see also \citealt{eggleton+2001_tides}). The fraction of stars in the halo with velocities above $\sim 700 $ km s$^{-1}$ decreased by $\sim 5\%$, but the distribution remains qualitatively similar and inconsistent with observations. Interestingly, tides decrease the collision probability by a factor of $\sim$2.\footnote{\citet{bradnick+2017} also observed a a decrease in the collision probability with tides.} We caution that a proper treatment of tides should use dynamic rather than equilibrium tides. We leave this for future work.

The outcome of the collision depends on relative impact velocity of the two stars. If this is less than the surface escape velocity, coalescence can occur \citep{ginsburg&loeb2006}. This will occur for most collisions of binary stars \citep{antoniniColl1+2010}.

\citet{antoniniColl2+2011} previously studied stellar collisions in tidally perturbed binaries with SPH simulations. They make several predictions for the appearance of merger remnants, which we summarize here.
Collision products will be rapidly spinning, and will initially be much brighter and larger than other stars of the same mass. However, they will rapidly contract to the main sequence after a thermal timescale ($\lsim 0.1$ Myr). 
Mergers also enhance the central hydrogen abundance, and create a younger-looking star. This star would stand out as a blue straggler in an older population. Although not as easily identifiable in a young population (e.g. the young Galactic Centre disc), merger remnants could be found via abundance anomalies in Lithium, Beryllium, and Boron.

Mergers may also account for the G2-like objects in the Galactic Centre \citep{prodan+2015}. G2 is a dusty gas cloud on an eccentric orbit ($e\sim0.97$) compact orbit ($a\sim 0.03$) \citep{gillessen+2012, gillessen+2013}. The cloud may contain a central star \citep{witzel+2014}, but this is disputed \citep{gillessen+2019}. In fact, a total of six G2-like objects have been identified Galactic Centre with a broad range eccentricities ($\sim0.1-0.97$), and inclinations ($\sim 52^\degree-118^\degree$ relative to the line-of-sight) \citep{ciurlo+2020}. An eccentric disc is capable of producing merger remnants with a broad range of orbital inclinations and eccentricities, but the predicted inclination distribution is inconsistent with observations. For example, in the disc simulations of \citet{gnm2021} particles that have passed within $3\times 10^{-4}$ pc of the MBH (``disruptees'' in \citealt{gnm2021}; these are the orbits that would produce collisions) are on orbits with inclinations of $\sim 12-170^\degree$ and eccentricities between $0.25$ and $0.998$ after $\sim 5$ Myr. In fact, 65\% of disruptees orbits are retrograde with respect to the disc, and their orbital planes are no longer correlated with it. This is inconsistent with the distribution of G2-like objects, which have prograde orbits with respect to the disc.

Merging stars could also bias the mass function of the disc and the S-stars, making them more top-heavy  \citep{prodan+2015}. 

\subsection{Centre-of-mass kicks}
We have neglected the back-reaction of binary encounters on the binary's centre-of-mass orbit. 
For example, if a binary expands due to tidal heating, the semimajor axis of its centre-of-mass orbit must shrink by energy conservation.

These kicks can shift the orbits of distance binaries substantially, moving them from the full to the empty loss cone regime.\footnote{For example, a 8 solar mass 0.1 AU binary at 100 pc can shift its outer orbit to $\sim$25 pc due to tidal expansion.} However, as discussed previously, the velocity distribution in both the full and empty loss cone regimes is inconsistent with the observations. 

Centre-of-mass kicks also change the velocity distribution of ejected stars directly, by pushing progenitor binaries into a deeper potential well.
In practice, this will always be a subdominant effect to the tidal expansion of the binary. The specific energy of the ejected star is

\begin{equation}
    \epsilon_{\rm ej} \approx \left(\frac{M_{\rm bh}}{m_{\rm bin}}\right)^{1/3} \epsilon_{\rm bin} \frac{2}{1+q} -\epsilon_{\rm com},
\end{equation}
where $\epsilon_{\rm bin}$ ($\epsilon_{\rm com}$) is the specific binding energy of the binary (centre-of-mass) orbit, and $q$ is the ratio between the ejected star's  mass and the other star's mass. This expression can be derived by Taylor-expanding the potential energy about the tidal radius (see also \citealt{kobayashi+2012}). If $\epsilon_{\rm bin}$ is perturbed, $\epsilon_{\rm com}$ will be perturbed by the same amount. However, the perturbation to the internal binary orbit will have a larger effect on ejected star's energy for $q\lsim 200$ due to the $\left(M_{\rm bh}/m_{\rm bin}\right)^{1/3}\approx 100$ term.

\subsection{Star formation history}
\label{sec:sfh}
So far, we have implicitly assumed constant star formation in all of our models. We now discuss the implications of deviating from this assumption. 

Considering the distances and masses of the observed sample, only the star formation history between a few$\times 10^7$ to a few$\times 10^8$ years ago is relevant. This means the inner few parsecs of the Galactic Centre are not a promising region for the producing the \citet{warren_brown+2018} sample, as there is little evidence for star formation there on these timescales \citep{pfuhl+2011}. Larger radii beyond $\sim$10 pc are more promising, as there is evidence of ongoing star formation \citep{nogueras-lara+2020}.

A star formation rate that increases towards the present day favors faster stars, and exacerbates the observational tension. On the other hand, a star formation rate that decreases towards the present day can relieve the tension. In particular, stars with velocities $\gsim 700$ km s$^{-1}$ (after deceleration) will travel beyond 140 kpc after $\sim 200$ Myr. Truncating star formation at this point will remove these stars. 

\begin{figure}
    \includegraphics[width=\columnwidth]{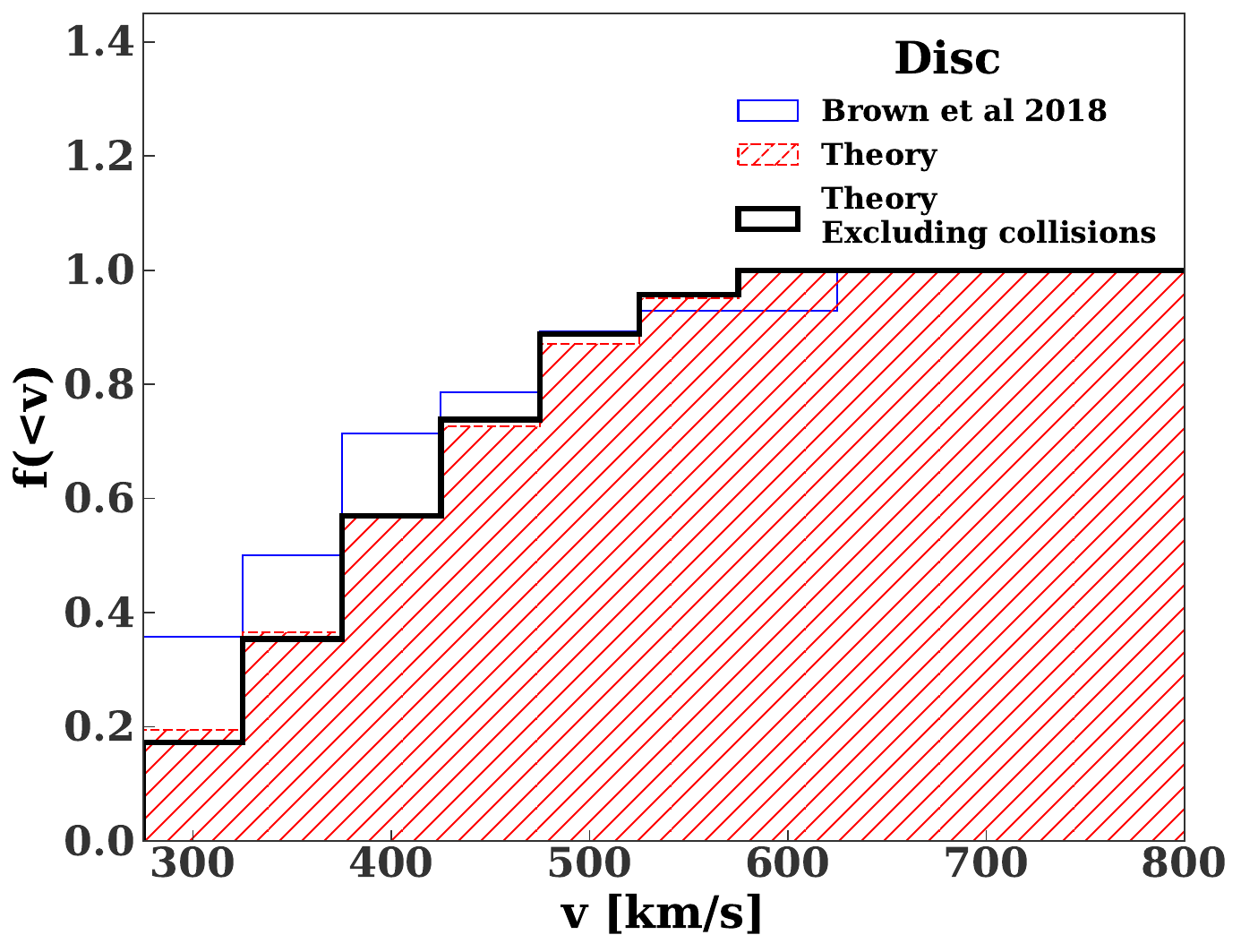}
    \caption{Same as Figure~\ref{fig:diskv}, except we exclude stars older than 200 Myr in the theoretical (red and black) distributions.}
    \label{fig:age_cut}
\end{figure}

We have tried excluding stars with ages less than 50, 100, and 200 Myr in our models.  After excluding stars with ages less than 200 Myr in our disc model, we can reproduce the observed velocity distribution in \citet{warren_brown+2018}, as shown in Figure~\ref{fig:age_cut}. (Here the KS-test p-value is 0.16).

\begin{figure}
    \includegraphics[width=\columnwidth]{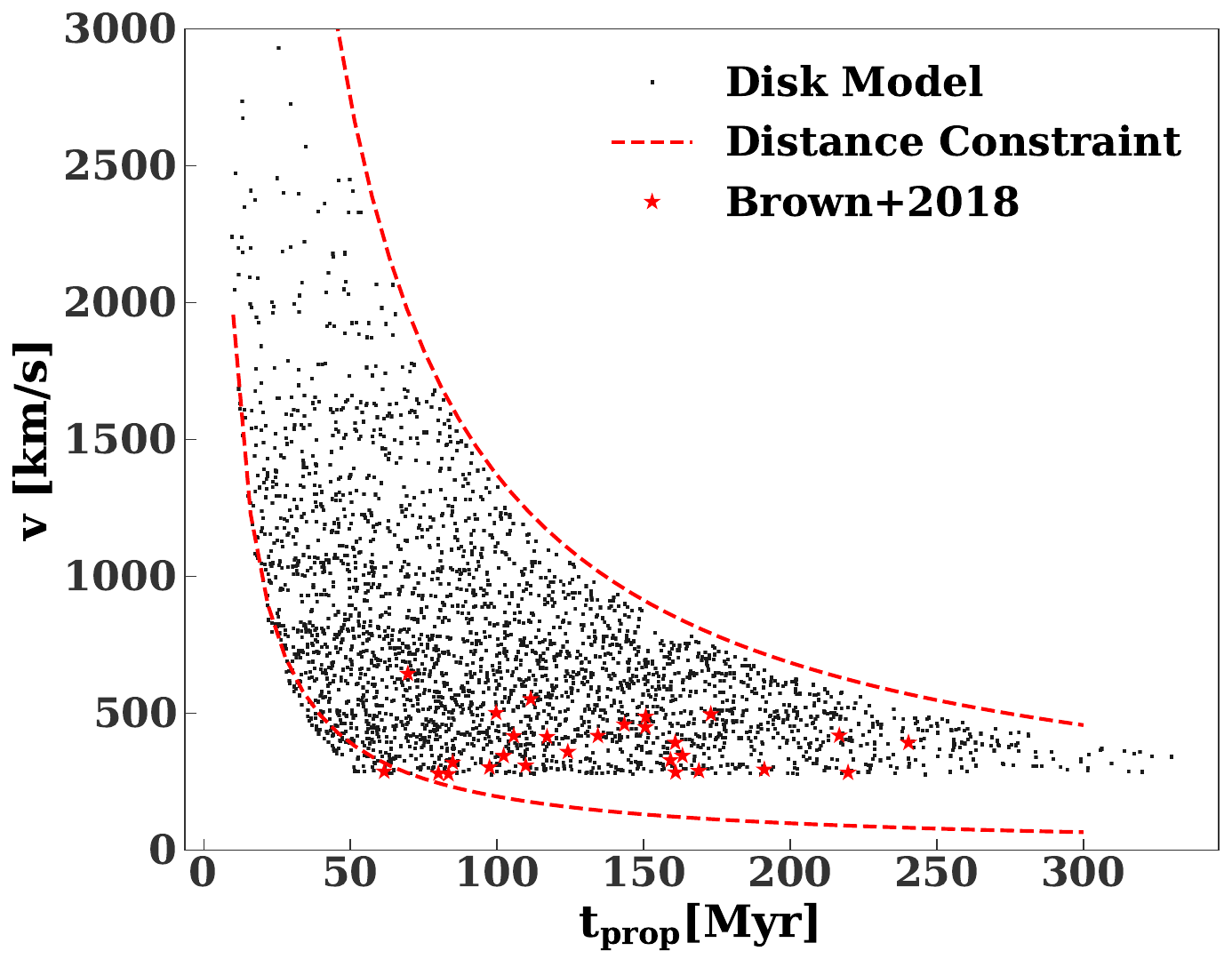}
    \caption{Black points show the velocity versus propagation time in the Galactic potential for our disc model, while the red stars show the observed distribution from \citet{warren_brown+2018} (assuming these stars actually come from the Galactic Centre). The dashed red lines show the minimum and maximum velocities consistent with a Galactocentric distance between 20 and 140 kpc.}
    \label{fig:tprop}
\end{figure}

This is not surprising. We have previously introduced cuts in Galactocentric distance. By introducing a minimum age, we are cutting off the high velocity tail of the distribution. This is illustrated in Figure~\ref{fig:tprop}, which shows the minimum and maximum velocity consistent with Galactocentric distances between 20 and 140 kpc as a function of propagation time. The black points are the velocities and propagation times from our disc model, and are bounded by these curves.\footnote{The curves are calculated assuming constant velocities. In reality, stars are decelerated in the Galactic potential, which results in a slight mismatch between the edges of the stellar distributions and the red curves.} By introducing a cutoff in the age, we are also introducing a cutoff in the velocity. 
The cutoff is sharpest and most extreme in the disc model, where the propagation time is equal to the stellar age. 

However, the required cutoff in age is ad-hoc, and would produce a truncation in the observed flight times at 200 Myr, which is not observed. In fact, the stars in the \citet{warren_brown+2018} sample have been traveling between $\sim 60$ and $240$ Myr (assuming they were ejected from the Galactic Centre), as shown by the red stars in Figure~\ref{fig:tprop}.

\subsection{Inclination changes and quasi-secular evolution}

In the inner spherical model we have neglected changes to the inclination of the binary's centre-of-mass orbit between close encounters. In fact the inclination should change due to both two-body relaxation and vector resonant relaxation (see \citealt{hamers+2018} and references therein). Changes in inclination may lead to the more chaotic evolution of binaries' eccentricity and inclination, instead of the ZLK-like oscillations we see in our simulations, which may increase the number of collisions (see the discussion in \citealt{bradnick+2017}). 

We have also neglected ZLK oscillations between encounters, which may lead to collisions between the binary stars or shrinking of the binary semimajor axis via tidal effects (see \citealt{antonini&perets2012,prodan+2015} and references therein). However, these oscillations may be detuned by kicks to the centre-of-mass orbit as discussed in \citet{bradnick+2017}.

Overall, we leave these effects for future work. While they may change the velocity distribution in the inner spherical model, it can likely be ruled out for other reasons. In particular, this model cannot realistically produce the observed hypervelocity stars considering (i) the observed star formation history in the Galactic Centre (see the discussion in \S~\ref{sec:sfh}) and (ii) the disruption rate required to produce the observed population would be unrealistically high.

\subsection{Observational data and tangential velocities}
Recently, \citet{kreuzer+2020} reanalyzed the sample in \citet{warren_brown+2018} finding somewhat different atmospheric parameters and spectro-photometric distances. Approximately 19  stars remain consistent with a Galactic Centre origin. 
We have repeated our observational comparisons with the radial velocities of these 19 stars from \citep{kreuzer+2020}, and find consistent results. This is not surprising, since the radial velocities in \citet{warren_brown+2018} and \citet{kreuzer+2020} are very similar, though in the latter work some stars have significantly higher total velocities. In particular, one star, HVS 22, has a velocity of $1530^{+690}_{-560}$\,km\,s${}^{-1}$, with substantial tangential components. 
However, tangential velocities cannot explain the tension with observations. Stars with velocities $\gsim 700$ km s$^{-1}$ in the halo should not show significant tangential velocities if they are ejected from the Galactic Centre \citep{yu&madau2007}.

\section{Summary}
\label{sec:summ}
Spectroscopic surveys have identified a few dozen unbound B stars in the Galactic halo over the last few decades. The origin of many of these stars in still unconstrained. In principle, they may originate from tidal disruption of binary stars in the Galactic Centre. However, we find the observed properties of these stars are inconsistent with this scenario. Our main results are summarized as follows:

\begin{enumerate}
\item We predict the velocity distribution of stars ejected by binary disruptions in the Galactic Centre for different loss cone filling mechanisms, using an empirically-motivated model for the binary population \citep{moe+2017}.

\item The observed velocity distribution is not consistent with expected distribution from binary disruptions. In particular, the observed velocity distribution is truncated above $\sim$700 km s$^{-1}$. This is difficult to account for in the context of binary disruptions in the Galactic Centre.

\item Binaries at small galactocentric radii can experience many close encounters with the MBH prior to disruption. This can truncate velocity distribution by causing binaries to expand in semimajor axis, or by causing disruptions to occur further from the MBH on average \citep{zhang+2010_hvs, zhang+2013_hvs}. However, this will only occur if the binaries undergo a slow diffusion in angular momentum, which implies a small disruption rate that is problematic for explaining the observed hypervelocity stars. Close-in binaries can be torqued to disruption more efficiently by a secular gravitational instability in a young stellar disc \citep{madigan+2009, generozov&madigan2020}. However, in this case binaries experience coherent changes in angular momentum over many orbits, rather than slow diffusion, and multiple encounters do not truncate the velocity distribution.

\item  In principle, repeated close encounters can also shift the velocity distribution downwards by causing the closest binaries to collide. In practice, we find collisions, although potentially frequent,  have a small effect on the velocity distribution of ejected stars in the halo. Depending on the model, the fraction of binaries that merge can vary between 1 and 50\% of the HVSs ejection rate.

\item Collisions can produce distinctive remnants with unusual abundance anomalies, and may account for the set of recently discovered dusty gas clouds in the Galactic Centre \citep{gillessen+2012, ciurlo+2020}. 

\item The star formation rate can be tuned to produce the observed velocity distribution in disc model. However, the required 
star formation history is inconsistent with the distribution flight times of the observed sample. 

\item Stars ejected from the Galactic Centre should leave counterparts in close orbits around its MBH. While there are young stars on close orbits there (the S-stars), their semimajor axis distribution is inconsistent with the observed velocity distribution of hypervelocity star candidates.

\end{enumerate}

Overall, these results suggest many of the unconstrained hypervelocity star candidates do not originate in the Galactic Centre, or that some unknown observational bias prevents detection of the fastest stars. 

Alternatively, the binaries in and near the Galactic Centre may have unusual distributions compared to the the samples in \citet{moe+2017}. For example there could be fewer close binaries than we assume.  However, the necessary adjustment is somewhat extreme. In order to reproduce the observed velocity distribution,  the semimajor axis distribution would have to be truncated at $\sim 0.3$ au, approximately an order of magnitude greater than the separation of a contact binary. 

\section*{Acknowledgements}
We thank the anonymous referee for helpful comments.

AG thanks Ann-Marie Madigan for helpful conversations on this work, and we thank angular momentum for keeping the binaries spinning.


\section*{Data Availability}

Data used in this study will be made available upon reasonable request.



\bibliographystyle{mnras}
\bibliography{master} 




\appendix

\section{Deceleration in the Galactic potential}
\label{app:decel}

Ejected stars are propagated through a model Galactic potential. In particular, we use the procedure in \citet{generozov2020}. Stars are first decelerated by the Milky Way's nuclear star cluster.\footnote{Here, we use the best fit density profile from \citep{schodel+2018} to compute this deceleration.}  

We then use the \texttt{gala} software package \citep{gala} to calculate the deceleration beyond 10 pc. In particular, we calculate stellar orbits in the default ``MilkyWayPotential'' class, which contains a Hernquist bulge and nucleus, a Miyamoto-Nagai disk, and an NFW dark matter halo. The parameters for each components are derived based on recent mass measurements between 10 pc and 150 kpc (the disk and bulge parameters are the same as those in \citealt{bovy2015}; the full list of mass measurements used to constrain the halo properties are drawn from various sources).\footnote{The full list can be found at \url{https://gala-astro.readthedocs.io/en/latest/potential/define-milky-way-model.html\#Introduction}.} The integration time is a random number between 0 and 500 Myr.


\bsp	
\label{lastpage}
\end{document}